# Known mechanisms that increase nuclear fusion rates in the solid state

Florian Metzler[1]*, Camden Hunt[2,3,4,5], Nicola Galvanetto[1,6]

[1]Massachusetts Institute of Technology, Cambridge, Massachusetts 02139, United States. [2]Department of Chemistry, The University of British Columbia, 2036 Main Mall, Vancouver, British Columbia, V6T 1Z1, Canada. [3]Stewart Blusson Quantum Matter Institute, The University of British Columbia, 2355 East Mall, Vancouver, British Columbia, V6T 1Z4, Canada. [4]Department of Chemical and Biological Engineering, The University of British Columbia, 2360 East Mall, Vancouver, British Columbia, V6T 1Z3, Canada. [5]Canadian Institute for Advanced Research (CIFAR), 661 University Avenue, Toronto, Ontario, M5G 1M1, Canada. [6]University of Zurich, Winterthurerstrasse 190, 8057 Zurich, Switzerland.

* corresponding author: fmetzler@mit.edu

We investigate known mechanisms for enhancing nuclear fusion rates at ambient temperatures and pressures in solid-state environments. In deuterium fusion, on which the paper is focused, an enhancement of >40 orders of magnitude would be needed to achieve observable fusion. We find that mechanisms for fusion rate enhancement up to 30 orders of magnitude each are known across the domains of atomic physics, nuclear physics, and quantum dynamics. Cascading such mechanisms could lead to an overall enhancement of 40 orders of magnitude and more. We present a roadmap with examples of how hypothesis-driven research could be conducted in—and across—each domain to probe the plausibility of technologically-relevant fusion in the solid state.

## 1. Introduction

Nuclear fusion is purported to require extreme temperatures or pressures[1–5]. Claims that nuclear fusion is achievable at ambient temperatures and pressures in solid-state materials have surfaced repeatedly, but were dismissed for lack of plausible explanations[6–10]. Readily observable deuterium (D–D) fusion at ambient conditions would require a fusion rate increase of >40 orders of magnitude compared to the baseline spontaneous fusion rate of deuterium gas (~$10^{-64}$/s per deuteron pair at the molecular distance of 74 pm[11]; in contrast, $10^{22}$ deuteron pairs in a ~3 g TiD$_2$ sam-

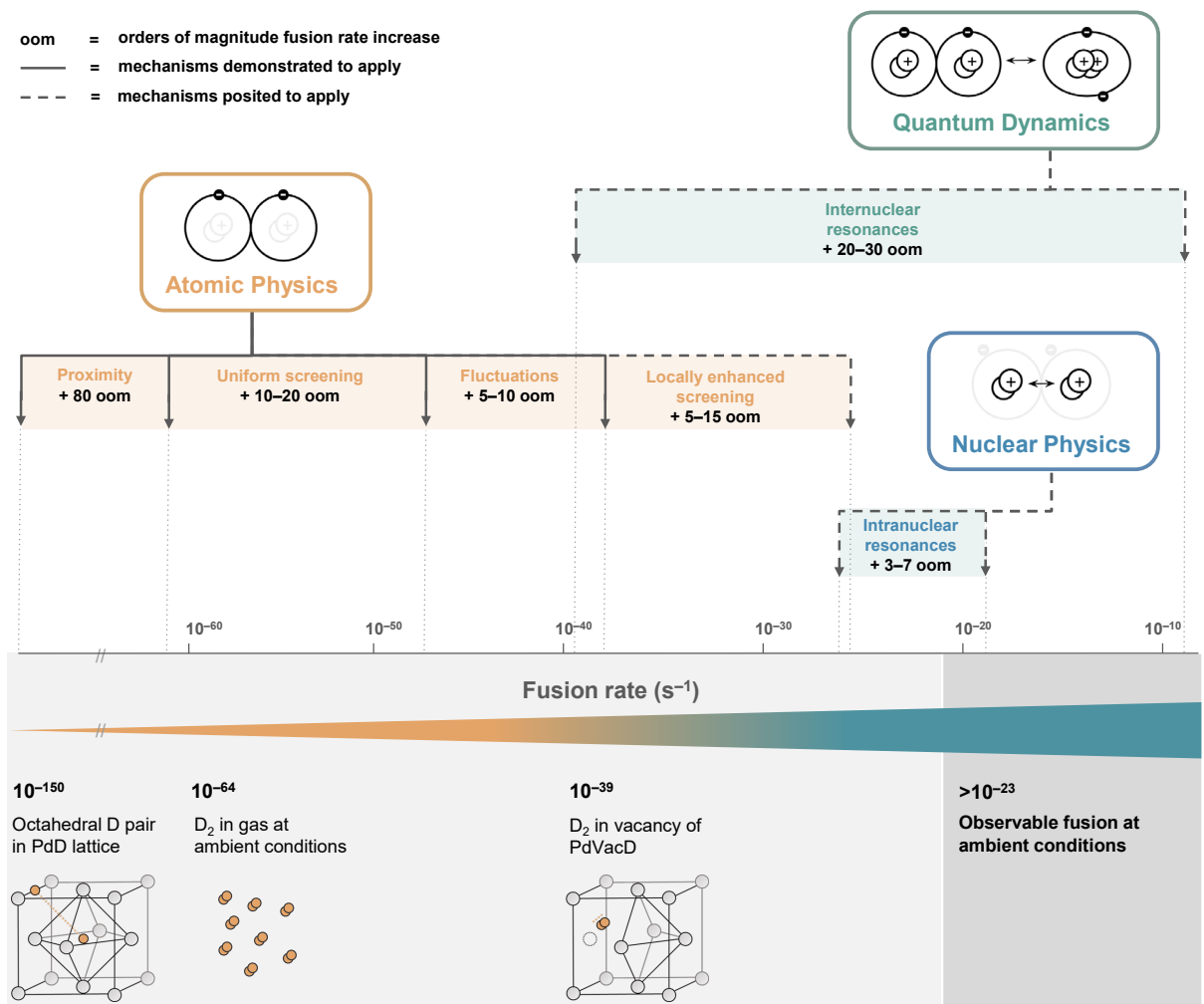

**Figure 1.** The three fields considered in this work and the estimated fusion rate enhancement that could be gained from singular mechanisms and combinations of mechanisms. A D–D fusion rate of $10^{-10}$/s would result in observable fusion in metal deuterides at ambient conditions (see Supplementary Note S0.1).

ple would exhibit 100 fusion reactions per second if the fusion rate were $10^{-20}$/s; see Supplementary Note S1.1). This paper examines D–D fusion in a solid-state environment through the perspectives of three disciplines — atomic physics, nuclear physics, and quantum dynamics — and asks the following question:

*What known mechanisms can increase nuclear fusion rates in the solid state?*

We offer no single mechanism that could increase fusion rates by 40 orders of magnitude. However, there are known mechanisms across fields that could individually increase fusion rates up to 30 orders of magnitude each (Figure 1). Identified mechanisms are discussed with supporting calculations.

The possibility of cascading these mechanisms provides a first-principles research map for studying solid-state fusion. We present a combination of enhancement mechanisms that could cumulatively provide a 50 orders of magnitude fusion rate increase (Figure 1). We conclude with examples of how hypothesis-driven research could be conducted in — and across — each field to probe the plausibility of technologically-relevant fusion in the solid state at ambient conditions.

## 2. The Atomic Physics Perspective

A simple description of nuclear fusion can be found in the Gamow model[12]. The Gamow model frames fusion as a quantum tunneling event through a potential barrier, with the barrier defined by the interatomic potential between two nuclei (Figure 2a). Here, nuclei are considered point masses and point charges, and the fusion event is treated as instantaneous. Accordingly, the Gamow model is concerned with atomic physics (the interatomic potential) rather than nuclear physics (the fusion event). The interatomic potential can be simple — such as with nuclei in a diffuse plasma — or complex — such as with nuclei in a dense solid, where the electronic structure of the solid will influence Gamow model variables as described below.

The interatomic potential (Figure 2a)[13–16] is defined by three partial potentials arising from electromagnetic and nuclear forces: i) the electrostatic repulsion between nuclei due to positive charges of protons; ii) the electrostatic attraction between such positive charges and negatively charged electrons as well as the electrostatic repulsion between such electrons; and iii) the short-range (<~10 fm) strong nuclear force attraction between all nucleons (see Supplementary Note S2.1). The resulting potential leads to an equilibrium position for any nuclei pair at the range of interatomic bond lengths (50–300 pm). If the interatomic distance of two nuclei is progressively decreased, the short-range attractive strong nuclear force will eventually dominate the electrostatic repulsion and the nuclei will fuse through quantum tunneling[17]. Quantum tunneling — a basic feature of all quantum systems — yields a probability distribution for overcoming the potential barrier as a function of the potential barrier and interatomic distance (Figure 2a). The product of the tunneling probability and the trial frequency (*i.e.*, the number of tunneling attempts) is the fusion rate. The trial frequency will manifest as collisons[18] (*e.g.*, in a plasma) or oscillations[19] (*e.g.*, in a solid). The two primary variables in Eq. (1) are the relative energy $E$ between the two nuclei and the size and shape of the potential barrier $V(r)$ between the two nuclei (Figure 2a). Most attempts at technologically-relevant fusion focus on maximizing $E$ (*e.g.* through heating to >100,000,000 K) instead of manipulating $V(r)$[1,20].

Situations with more than two interacting atoms — such as deuterons within a palladium lattice — introduce new considerations to the Gamow model. $V(r)$ is now a function of all interacting nuclei and electrons[21]. Moreover, the interatomic distance $(r_2-r_1)$ is now dictated by where the deuterons reside in the lattice. When the tunneling probability is evaluated over a range of interatomic distances, the importance of this latter point is apparent (Figure 2b-c). For example, the D–D fusion rate predicted by the Gamow model for face-centered cubic (FCC) palladium deuteride ($PdD_x$) has a range of ~60 orders of magnitude depending on deuteron occupancy (see Supplementary Note S2.2). If only octahedral sites are occupied, the fusion rate would be $10^{-110}$/s. This rate is further increased to $10^{-100}$/s if both octahedral and tetrahedral sites are occupied. This would be further increased to $10^{-64}$/s if vacancies in the lattice capable of accommodating $D_2$ are occupied (before considering screening effects as discussed below). It is empirically demonstrated for $PdD_x$ alloys that site occupancy is a function of deuterium content, with a low (x < 0.1) deuterium content favoring octahedral site occupation[22], a medium (x > 0.1) deuterium content favoring octahedral and tetrahedral site occupation[23], and a high concentration (x > 0.9) purported to favor vacancy formation[24–27]. Recent computational studies suggest that many $M(H/D)_x$ alloys can form vacancies in the lattice capable of accommodating $H_2/D_2$[28,29].

These considerations extend beyond proximity. The free electron density in Pd is ~$6.8 \times 10^{22}$/cm$^3$, which corresponds to approximately one free electron per lattice atom[30]. The potential barrier between nearest-neighbor deuterons in a palladium-deuterium ($PdD_x$) alloy will be modified by this free electron density — a process known as electron screening[31]. Simplistically, electron screening reduces the positive electrostatic repulsion between nuclei and results in a reduced interatomic distance $(r_2-r_1)$ and is expressed as a reduction of $V(r)$ by a constant screening value $(U_e)$[31–35]. The two electrons in gas-phase $D_2$ correspond to a theoretical $U_e$ value of approximately 20 eV[36,37]. Theoretical values of Ue range from 50–150 eV for deuteron pairs in different metals[38,39]. A $U_e$ of 150 eV would correspond to a bond length reduction from 74 pm to 57 pm[40]. We calculate the effect of screening energies across this range (see Supplementary Note S2.3) for D–D fusion and find that it could provide an enhancement of fusion rates for $D_2$ upwards of 20 orders of magnitude in comparison to gas phase $D_2$ (Figure 2c). This is qualitatively consistent with experimental observations from accelerator experiments[34,38,41]. However, quantitatively, experimental observations of fusion rates in solids appear to exceed theoretical estimates of $U_e$[34,42]. Such discrepancies could be explained by local concentrations of electron density in the lattice ("locally enhanced screening" in Fig. 1), by dynamical increase of proximity and screening ("fluctuations" in Fig. 1), or by other enhancement mechanisms beyond the domain of atomic physics, as discussed in the following sections ("internuclear resonances" and "intranuclear resonances" in Fig. 1).

Lattice imperfections such as impurities and defects lead to local changes in the electron band structure, which can be interpreted as a larger curvature of the electron valence band and thus as a larger effective electron mass m$^*$[43]. Since the screening potential $U_e$ depends on the effective electron mass m$^*$ (see Supplementary Note S2.3), some lattice sites would then provide for higher screening. Czerski et al. 2020[44], referring to Zhang et al. 1995[45], argue that the effective electron mass m$^*$ can locally be larger than the electron rest mass $m_e$ by a factor of 9, corresponding to an increase of $U_e$ by a factor of 3. This conjecture is consistent with accelerator experiments that suggest a $U_e$ of 300 eV from the deuteron bombardment of vacancy-rich Zr, whereby the theoretically predicted $U_e$ of pristine Zr is 112 eV[46].

The discussion so far only considers equilibrium conditions. In actuality, deuterium nuclei inside a metal lattice are dynamic[47–50]. The timescale discrepancy of atomic/electronic oscillations (>>1 fs) and fusion events (<<1 fs) is large. This discrepancy implies that short-lived extrema in atomic position and electron density are effectively infinitely long from the perspective of two fusing nuclei. To account for dynamic atomic po-sitions, the tunneling probability must be evaluated across all occurring proximities resulting from fluctuations[51]. Fluctuations in proximity are caused by factors such as lattice oscillations (phonons) or atomic diffusion and are on the order of ± 5 pm[51]. This change in interatomic distance corresponds to an expected enhancement of the D–D fusion rate of ~8 orders of magnitude (see Supplementary Note S2.4). Fluctuations in electron density are also relevant; it was recently proposed that the use of electronic fluctuations (plasmons) could temporarily increase electron density between deuterons in $PdD_x$, with a calculated increase in $U_e$ of 40%[32,33]. This increase corresponds to an expected enhancement of D–D fusion rates of ~15 orders of magnitude.

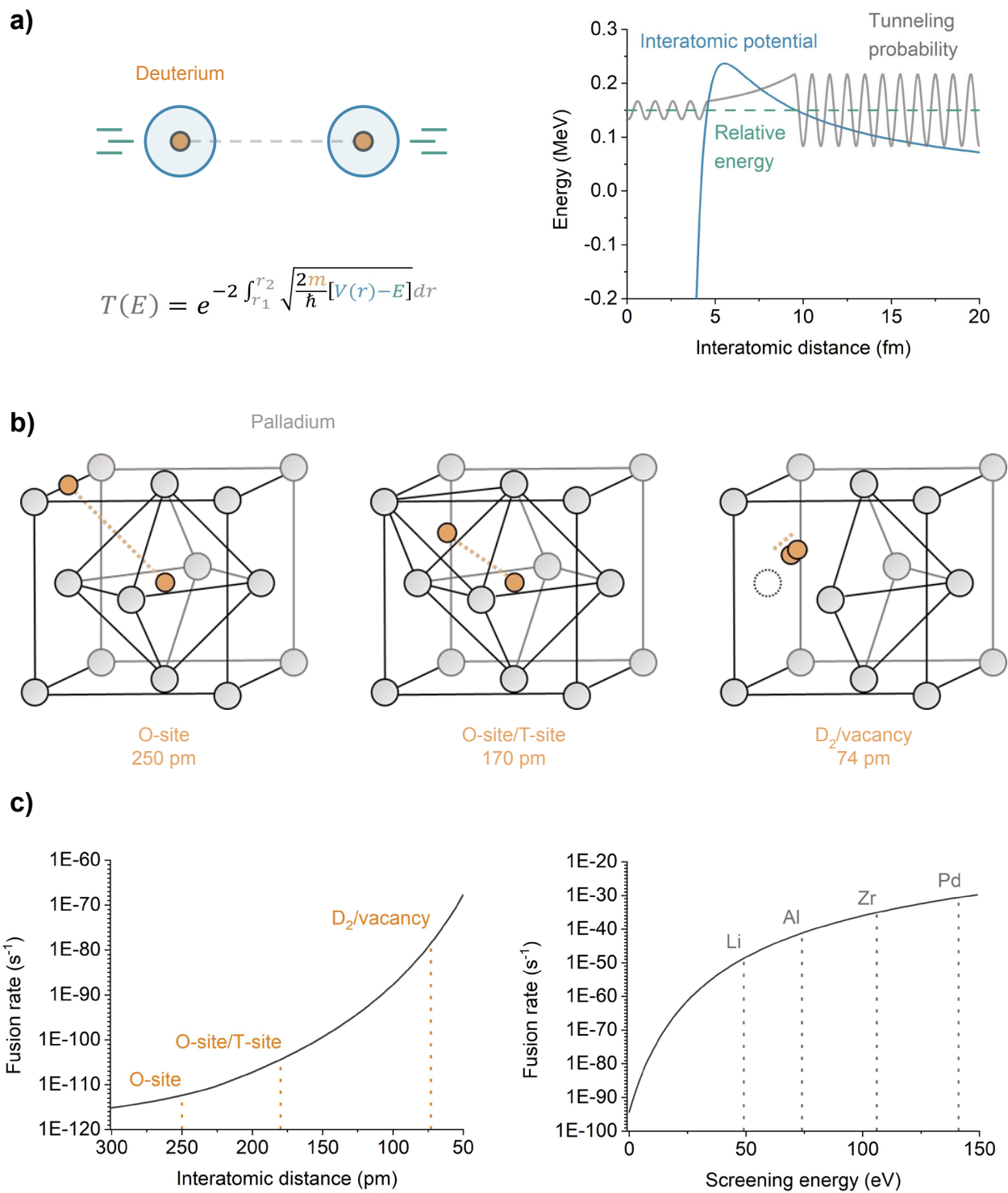

**Figure 2. a)** Application of the Gamow model for describing fusion of two deuterons (orange). $r_2-r_1$ denotes the interatomic distance, $m$ denotes the mass, $V(r)$ denotes the interatomic potential, and $E$ denotes the center-of-mass energy. **b)** Deuteron proximity as a function of lattice site occupancy. **c)** The effect of proximity and screening on D–D fusion using Gamow model assumptions. All calculations with detailed descriptions can be found in the Supplementary Note S1.

The atomic physics perspective contains mechanisms that will — non-contentiously and demonstrably — change fusion rates in the solid state: the effect of proximity as determined by equilibrium positions of hydrogen nuclei in their respective environments (site occupation); increased proximity caused by interactions of hydrogen nuclei with free electrons (electron screening), and; further temporal enhancement of proximity (structural and electronic oscillations. These mechanisms manifest to differing degrees in any metal-deuterium system. Our evaluation suggests a fusion rate enhancement upwards of ~25 orders of magnitude can arise from such mechanisms without controversy. Application of this enhancement to a fusion rate of $10^{-64}$/s for $D_2$ at ambient temperature and pressure yields a fusion rate of $10^{-39}$/s. This is ~20 orders of magnitude short of the alleged "cold fusion" rate of $10^{-23}$/s. Some authors have argued that locally enhanced screening effects in the vicinity of lattice defect sites may further increase fusion rates by up to 15 orders of magnitude[44]. However, more research is needed to substantiate such conjectures. In either case, observable fusion at ambient conditions — when viewed exclusively through the lens of both atomic physics — is not achievable.

## 3. The Nuclear Physics Perspective

Nuclear physics frames fusion with more granularity than atomic physics. Here, nuclei are seen not as point masses and point charges but as objects with intranuclear structure[52]. The need for this increased granularity is apparent from empirical fusion data[53] (Figure 3a). For example, proton-boron (p–[11]B) fusion exhibits peaks of the reaction probability near 162 keV and 675 keV. Atomic physics alone cannot account for these peaks. The Gamow equation yields a smooth curve when reaction probabilities are evaluated as a function of energy[12]. Differences between reaction probabilities predicted by the Gamow model and empirical reaction probabilities are attributed to intranuclear structure that results from nucleon interactions[54,55]. For p–[11]B fusion, the observed peaks originate from the resonant dynamics of the resulting [8]Be[4]He nucleon cluster — a temporary molecule-like configuration of the twelve nucleons — which then rapidly disintegrates through sequential alpha decay into three [4]He nuclei[56]. For deuterium-tritium (D–T) fusion, a broad peak centered around 100 keV is attributed to a resonance associated with the polarized $3/2^+$ spin state of the resulting [5]He nucleus, which then disintegrates into a [4]He nucleus and a neutron[57]. Resonances – i.e. increased reaction probabilities at specific energies — occur when a particular configuration of nucleons can favorably accommodate a discrete amount of energy (Figure 3b). Such conditions enable higher fusion rates if the energy imparted to the reactants is well-matched to their collective dynamics[58–61].

The p–[11]B and D–T resonances were not predicted, but measured. Predicting resonances using first-principles nuclear models is still aspirational due to the complexity of calculating nucleon-nucleon interactions[54,55,62–64]. This complexity can be divided into three challenges: i) the strong nuclear force exhibits three-nucleon effects in addition to nucleon-nucleon effects; ii) the strong nuclear force is strongly attractive at 1 fm < 2 fm, yet strongly repulsive at <1 fm, and; iii) the strong nuclear force is not organized around a center. Classical computers cannot easily handle these challenges[65]. Accordingly, our understanding of intranuclear resonances is presently limited to what is experimentally accessible[53,62,63,66].

No resonance has been experimentally detected for D–D fusion in the well-characterized high-energy (10–1000 keV) range[53] (Figure 3a). The medium-energy (5–10 keV) range is less characterized for D–D fusion, but the sparse data reported suggests a fusion rate substantially higher than what is predicted by atomic physics[34,36,39]. A hypothesis explaining unexpectedly high D–D fusion yields in the medium-energy range is the existence of a resonance for the resulting [4]He nucleus close to the reaction threshold at 23.84 MeV[46,72]. The maximum of this resonance would be in the low-energy range (<5 keV), with a tail that extends into the medium-energy range[41]. Knowledge of such a resonance could enable tuning of reactant energies to increase D–D fusion rates at low energy. This hypothesis presents a clearly-defined issue, as there is limited published data for D–D fusion — or any fusion process — in the low-energy range (Figure 3a). Moreover, the sparse data sets in this range are inconsistent between experiments for purportedly identical materials. Reliable <5 keV fusion rate data is necessary to develop low-energy fusion models required to verify or disprove the [4]He narrow-resonance hypothesis — and any other hypotheses related to the role of intranuclear structure at low and medium energies. The sparsity of such data has a prosaic explanation that requires understanding how fusion rate measurements are performed and how reaction probabilities are calculated from the experimental data.

Inconsistent data in prior work has been attributed to experimental challenges[41]. Fusion rates are measured by accelerating particles (e.g., $H^+$, $D^+$) into collision targets containing reactants (e.g., D, T, [11]B) and then detecting the resulting nuclear products (e.g., $y$, $n$, [4]$He^+$, $H^+$). The fusion cross section, $\sigma(\bar{E})$ — which represents the fusion probability — is subsequently determined by solving the thick-target yield equation (Figure 3c). Determination of $\sigma(\bar{E})$ requires measuring four variables: $d\bar{E}/dx$ (the stopping power of incident particles, which relates to the range); $Y_t$ (the yield of nuclear products); $N_d$ (the nuclear reactant density), and $f(E)$ (the center-of-mass energy of incident particles). There are fundamental challenges for determining each variable for low (<5 keV) energy experiments. $d\bar{E}/dx$ is on the same scale as most material passivation layers (<100 nm)[38]. Due to differences in materials processing, the passivation layer composition and thickness will vary between targets that are presumed to be identical[36,67]. This implies that low-energy fusion experiments conducted by different research groups that appear identical may not even be performed on the same material. Moreover, measurements of $Y_t$ for low-energy experiments are weak enough to challenge the signal-to-noise limits of modern nuclear diagnostics[35,38,68]. Low observable fusion rates necessitate long experiments, which introduces the additional issue of $N_d$ changing with time[36,64]. Finally, many experimental setups measure a wide distribution of $f(E)$ values that make it difficult to identify narrow resonances[35,38]. The consequence of these issues is sparse — and inconsistent — low-energy fusion rate data. If the role of intranuclear structure for low-energy fusion is to be clarified, experimental apparatus that enable control of the variables within the thick target yield equation must be developed.

Nuclear physics describes intranuclear resonances that demonstrably increase fusion rates at high energy (10–1000 keV)[53]. Similar resonances at low energy (<5 keV) are hypothesized to account for unexpectedly high fusion probabilities observed at medium energy (5–10 keV) that exceed values predicted by atomic physics[34,36,39]. Due to the complexity of nuclear modeling, such hypotheses are presently only testable through the acquisition of low-energy fusion rate data[69]. Collecting reliable data sets at low energies is an unsolved challenge in experimental nuclear physics[35,68]. Until this challenge is met, ambiguity about the role of intranuclear structure

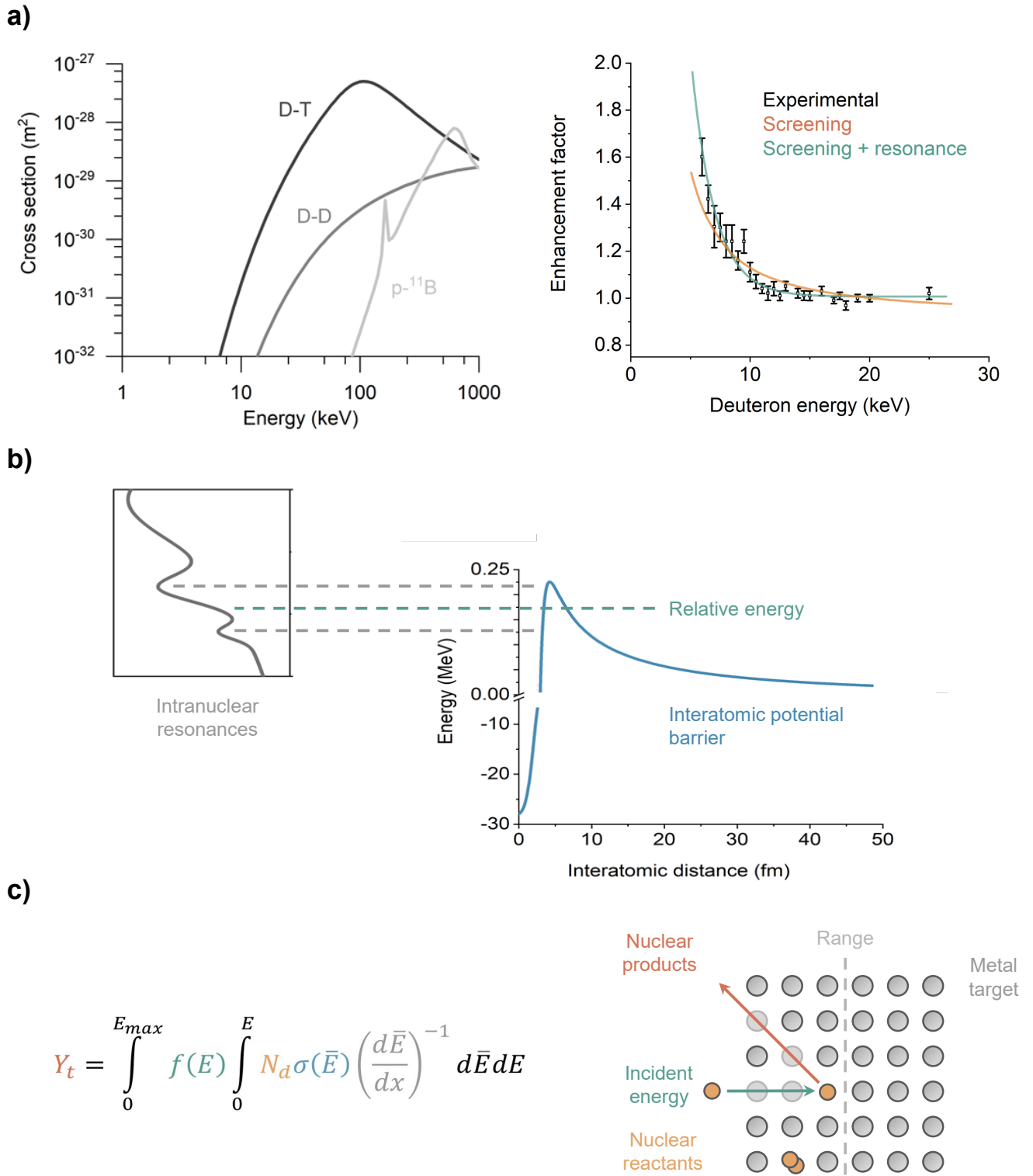

**Figure 3. a)** Empirical fusion probabilities of different fusion reactions (left) and deuteron fusion rate data (right). **b)** A pictorial representation of how intranuclear resonances can cause certain incident energies to exhibit higher tunneling probabilities than predicted using Gamow model assumptions. **c)** The thick-target yield equation (left) and a graphical interpretation of relevant variables (right) used to determine empirical cross-sections in solid-state targets. $Y_t$ is the yield of nuclear products, $f(E)$ is the center-of-mass energy of incident particles, $\sigma(\bar{E})$ is the cross section, $N_d$ is the nuclear reactant density, and $d\bar{E}/dx$ is the stopping power of incident particles.

in low-energy fusion will remain.

However, there is a large body of work that aims to predict nuclear resonances[46,57,64,66,70,71]. This includes a four-nucleon resonance that would explain unexpectedly high fusion rates measured in the medium-energy range. A combination of theoretical considerations and comparisons with experimental data led Czerski 2022[72] to predict resonance-based D–D fusion rate enhancements by 3-7 orders of magnitude in the eV range (see Supplementary Note S3 for a discussion of this predicted resonance and Fig. 3a for the experimental data used to motivate and calibrate it). Meaningful utilization of such a resonance would then require a high degree of precision and control over reactant energies and sample conditions. Application of this enhancement mechanism to a fusion rate of $10^{-39}$/s for $D_2$ in a palladium lattice with uniform electron screening at ambient conditions yields a fusion rate of $10^{-32}$/s. When combined with the 15 orders of magnitude enhancement from locally increased screening at lattice defect sites, as also posited in Czerski 2022[72], an overall fusion rate upwards of $10^{-19}$/s results. This is positioned within range of alleged "cold fusion" rates of $>10^{-23}$/s. Observable fusion at ambient conditions — when viewed through the lens of both atomic physics and nuclear physics — appears conceivable.

## 4. The Quantum Dynamics Perspective

Quantum dynamics — the study of the evolution of quantum systems as a function of time — frames nuclear fusion distinctly from atomic physics and nuclear physics. Here, fusion is treated as a quantum state transition rather than a quantum tunneling event. From this viewpoint, a deuteron pair $D_2$ in close proximity is an excited state of $^4He$[73,74]. The D–D reaction is represented by the $D_2$ four-nucleon cluster (i.e., the excited state) relaxing down to the more compact $^4He$ four-nucleon cluster (i.e., the ground state). The fusion rate is then equivalent to the state transition rate (Figure 4a) — and increasing the deexcitation rate corresponds to increasing the fusion rate.

Mechanisms for enhancing state transitions rates have been experimentally demonstrated at both the atomic[75–79] and the nuclear level[80–84]. This acceleration occurs if the excited atom or nucleus is provided with alternative channels for deexcitation which are faster than conventional channels. The fastest channels typically involve quantum coherence across interacting atoms or nuclei[85,86].

Two classes of coherence-related phenomena promote such accelerations: The first one is resonance energy transfer (RET)[58,87], a mechanism in which excitation energy is transferred nonradiatively from one system to another system that is coupled. This means that the transfer of energy is achieved not through the emission of a particle carrying the excitation energy, but through the coherent coupling of both systems to a common field. The electromagnetic field is an example of such a common field. At the atomic level, this mechanism is widely known and exploited in biology and chemistry (FRET experiments)[88]. Nonradiative energy transfer is conventionally believed to play a much less prominent role at the nuclear level because of the weakness of couplings that affect nuclear energetic states, compared to their much larger state transition energies. However, these weak coupling strengths can be increased under particular circumstances (see Box). Nonradiative transfer of nuclear energy has been theoretically described and experimentally emulated with precisely engineered resonance cavities[84,89–93].

The second mechanism to accelerate deexcitation is superradiance[94]. This is a phenomenon in which $N$ excited emitters are coupled together, which then deexcite at a rate proportional to $N^2$—which is much faster than the typical exponential decay[82,83]. In recent experiments, multiple nuclei were collectively excited by an X-ray laser[80]. The coupled nuclei comprised a quantum system in superposition that was able to absorb and emit nuclear excitation as a single coherent system. An acceleration of the deexcitation of the first excited state of $^{57}Fe$ nuclei proportional to the number of excited nuclei has been experimentally observed this way[80] —as predicted in 1954 by Dicke[94] and, for nuclei, in 1965 by Terhune & Baldwin[95]. These studies empirically demonstrate that nuclear decay can be manipulated through quantum coherent effects in the solid state. Exploratory experiments with $^{57}Fe$ nuclei suggest that phonon-mediated coupling may result in transition rate increases of several orders of magnitude for $^{57}Fe$ excited states[81].

Mechanisms of resonance energy transfer and superradiance-like enhancement can be combined: When a large number of atoms $N$ or nuclei participate in a coherently coupled quantum system within which RET occurs, then transfer can be accelerated by an enhancement factor up to $N^2$. Coherently accelerated RET is referred to as *supertransfer* and has been experimentally demonstrated at the atomic level[96,97]. Supertransfer is expected to also manifest in excitation transfer at the nuclear level[93].

An increased deuterium fusion rate for molecular $D_2$ in a metal lattice that results from accelerated nuclear state transitions is conceivable if: i) a coupling – even if weak – exists between nuclear states of nearby nuclei; ii) this coupling produces a collective state across many participating nuclei in the coherence domain; and iii) coupled acceptor nuclei exhibit states that are well-matched to the energy held by the $D_2$ donor state (23.84 MeV). A precise match for the $D_2$ donor state is a coupled $^4He$ nucleus, as it is capable of accommodating the energy released in the $|D_2\rangle \rightarrow |^4He\rangle$ transition through the complementary $|^4He\rangle \rightarrow |D_2\rangle$ transition. In the presence of coupling and supertransfer enhancement (Figure 4b), the accelerated deexcitation by enhanced resonance energy transfer $|D_2\ ^4He\rangle \rightarrow |^4He\ D_2\rangle$ is conceivable (whereby the resulting state can be either localized or delocalized). This process would manifest as oscillations between two population states, i.e., Rabi oscillations. The transfer rate for this process is estimated to be on the order of $10^{-34}/s$[98] – which would be a ~30 orders of magnitude faster than the $|D_2\rangle \rightarrow |^4He\rangle$ spontaneous emission rate of $10^{-64}/s$ (i.e. the tunneling probability in the Gamow model) – see Supplementary Note S4.1.

If energy redistribution remains limited within a coherent closed quantum system, then it is experimentally difficult to be detected and practically not exploitable. In the example above, every accelerated (exothermic) fusion event $|D_2\rangle \rightarrow |^4He\rangle$ would be offset by a matching (endothermic) fission event $|^4He\rangle \rightarrow |D_2\rangle$ with all energy remaining within the closed system. However, if acceptor nuclei incoherently disintegrate before the next transfer, the closed quantum system becomes open, with energy leaving in the form of energetic particles. In a metal-deuterium lattice with impurities, many common nuclei exhibit a dense number of excited states in the MeV range and represent acceptor candidates for nuclear resonance energy transfer processes (see Supplementary Note S4.2). Upon resonantly receiving a large quantity of energy, many nuclei would disintegrate through nuclear emission processes ($^4He$, $H^+$, $n$, $\gamma$). The result would be an increase in observed fusion products and a variety of different fusion products based on the identity of the impurities – as is suggested by earlier experimental reports (see Supplementary Note S5)[99–102]. This implies that rigorous fusion research in solid-state materials may necessitate purity requirements as well as deliberate doping similar to the semiconductor industry[103].

If acceptor nuclei are present in the coupled system that can partially absorb the donor energy, resonance energy transfer to multiple acceptor nuclei becomes possible, as long as the donor energy matches the total acceptor energies[104,105] (Figure 4c). The conversion of large energy quanta into multiple smaller quanta is known as downconversion[104,105].

Quantum dynamics offers mechanisms for increasing state transition rates at both the atomic and nuclear levels. Acceleration of atomic[75–79] and nuclear[80–83] state transitions has been empirically-demonstrated. Treatment of nuclear fusion as a quantum state transition provides underexplored handles for fusion rate enhancement. Resonance energy transfer mediated by interactions between nuclear states and oscillator modes available in the lattice, enhanced by superradiance effects are candidate phenomena for increasing fusion rates in the solid state. For the situations considered above — $D_2$ coherently coupled to resonant acceptor nuclei (Figure 4b and 4c) — we calculate the relative change in the D–D fusion rate and find an increase upwards of 30 orders of magnitude (see Supplementary Note S4.2). Application of this enhancement to the baseline spontaneous fusion rate of $10^{-64}$/s at ambient conditions yields a fusion rate of $10^{-34}$/s. Adding a screening potential of 150 eV to the calculation, per the discussion in section 2, yields a fusion rate of $10^{-18}$/s. This is positioned well within the range of the alleged "cold fusion" rate of $10^{-23}$/s. Observable fusion at ambient conditions — when viewed through the lens of both atomic physics and quantum dynamics — is conceivable.

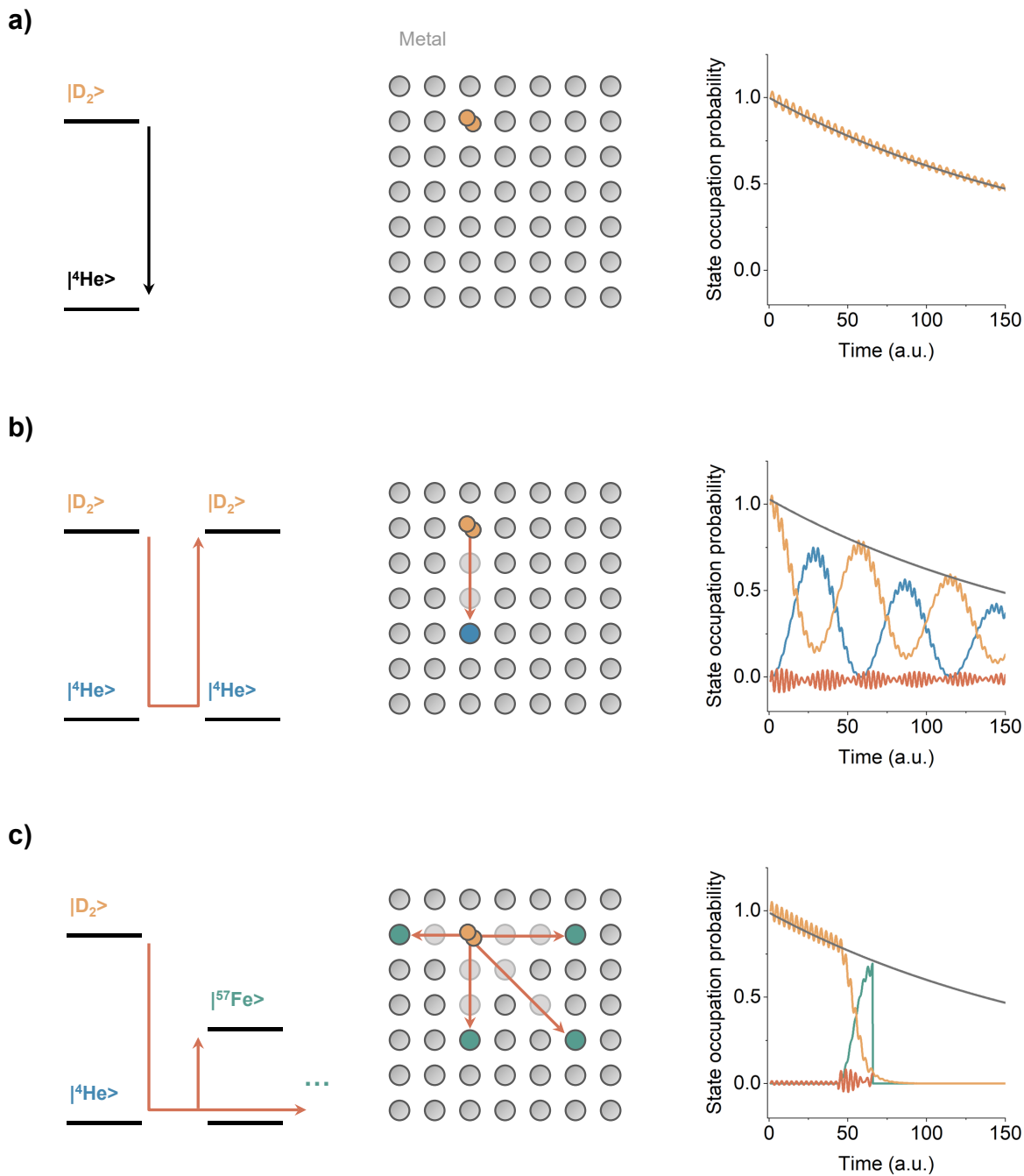

**Figure 4. a)** Nuclear fusion as spontaneous emission of a two-level system between an excited state ($D_2$) and ground state ($^4$He). The result is an exponential probability distribution analogous to the Gamow model. **b)** Resonant energy transfer between coupled $D_2$ and $^4$He. $D_2$ and $^4$He exhibit perfect resonance that results in no energy leaving the system, and the energy transfer shows no observable effects. **c)** Downconversion of $D_2$ to $^4$He through matching states of nearby nuclei (e.g. $^{57}$Fe). Disintegration results in energy leaving the system, and the energy transfer shows observable effects (i.e. an increase in fusion rate and a change in the nuclear product distribution). Models representing these systems are discussed in the Supplementary Note S3.

> ### Interactions between nuclei
>
> Quantum systems such as atoms or nuclei can interact with one another through fundamental forces, thereby forming superposition states that can redistribute energy. Interactions can be mediated by oscillators, which are treated as shared quantum fields. The extent to which such interactions manifest (i.e. the coupling strength) determines the rates of induced dynamics (*e.g.*, $^{57}Fe^* \rightarrow {}^{57}Fe + \gamma$).
>
> Interactions between nuclei are canonically weak compared to interactions between atoms. At the atomic level, dominant interactions originate from the antenna-like nature of atomic dipoles. Comparable interactions exist at the nuclear level but are weaker due to the small dipole of nuclei. Electromagnetic interactions between atoms can be on the order of 50 meV (*e.g.*, photosynthesis[106,107]) and between nuclei they are on the order of 0.1 neV (*e.g.*, nuclear magnetic resonance[108,109]).
>
> Both nuclear and atomic interactions can be driven by external stimulation. For example, at the atomic level, phonon-qubit coupling can be increased by expanding the phonon population in a phononic cavity[101]. At the nuclear level, magnon-nuclear coupling can be increased by an external radio frequency (RF) field and by internal magnetic fields, whereby the induced coupling scales with the field strength[111,112]. Internal magnetic fields can be activated via the stimulation of phonons and spin waves[113]. For both phonon-qubit and magnon-nuclear systems, coupling strengths of 100 neV are typical[77,111]. Some examples for undriven and driven couplings at the atomic level and the nuclear level are shown below:
>
> |  | Undriven | Driven |
> | --- | --- | --- |
> | Atomic Level | Electric dipole-dipole coupling between molecules (*e.g.*, between chlorophylls in photosynthesis). Coupling strength: ~50 meV[114] | Phonon-qubit coupling (*e.g.*, a phononic cavity as a quantum bus between superconducting circuits). Coupling strength: ~100 neV[110] |
> | Nuclear Level | Magnetic dipole-dipole coupling between nuclei (*e.g.*, between between C nuclei in $^{13}C$ NMR measurements). Coupling strength: <0.1 neV[115] | Magnon-nuclear coupling (*e.g.*, $^{57}Fe$ nuclei interacting with an RF field). Coupling strength: ~100 neV[111] |
>
> Phonon-nuclear coupling is suggested to be larger still and also externally driven[116]. The coupling originates from the center-of-mass contribution to nucleon motion resulting from lattice oscillations[116]. This contribution represents a boost to the nuclear spin-orbit coupling, which affects nuclear states. Coupling strength estimates for $^{181}Ta$ nuclei interacting with a 10 THz field are on the order of meV[117].

## 5. Conclusion

Within the three major communities presented — atomic physics, nuclear physics, and quantum dynamics — mechanisms are known that increase nuclear fusion rates in solid-state materials. No individual mechanism can traverse the ~50 orders of magnitude necessary to make solid-state fusion at ambient conditions technologically relevant. A combination of them might.

This final claim is only useful if it leads to testable hypotheses. Consider D–D fusion in $PdD_x$ through the perspective of the three fields discussed above. Considerations from the atomic physics perspective necessitate that $D_2$ exists in the lattice. This insight provides a clear research target to pursue for fusion rate studies — vacancy-rich $PdVac_yD_x$ that can accommodate $D_2$. This target can be pursued without contention, as vacancy-rich metal-hydrogen alloys are of interest to multiple communities. Nuclear physics suggests that anomalously high D–D fusion rates at low energy (<5 keV) can be explained by intranuclear resonances centered in the <1 keV range. D–D fusion rate measurements in that energy range would provide clarification. The challenges associated with collection of such data are firmly in the domain of materials science and nuclear diagnostics, and would benefit from collaboration between chemists and physicists. This gives a clear research target to pursue — the development of experimental apparatus for the collection of high resolution D–D fusion data in the <5 keV range. Again, this pursuit advances our understanding of nuclear physics and can be done without contention, as the data would be invaluable for nuclear model development. Quantum dynamics suggests possible mechanisms of significantly increasing nuclear reaction rates in the solid state through quantum coherent effects. We should not disregard the possibility of accelerating fusion in the solid state using coherent coupling, particularly since coupling mechanisms have been empirically demonstrated to accelerate other nuclear state transitions. This gives a clear research target to pursue — fabrication of materials with controlled quantities of resonant dopants (i.e. acceptor nuclei) and subsequent study of nuclear transformations in these materials under coherent stimuli (e.g. phonons). Such studies could ideally be expanded to include vacancy-rich target materials with high hydrogen loading that also benefit from proximity and screening enhancements (e.g. $PdVac_yD_x$ with x > 0.9 and y > 0.01). Experiments reporting unexpected nuclear products from low-energy stimulation of metal-hydrogen samples[99-102,118,119] in principle could fit this criteria, but well-controlled investigation of such systems are absent. Observable fusion at ambient conditions — when viewed through the lens of atomic physics, nuclear physics, and quantum electrodynamics — is a research subject of interest.

The research targets above cross multiple scientific disciplines. Cold fusion research requires no sacrifice when it is recognized as a question that will be asked and answered through the advancement of established fields. We call on the scientific community to explore the mechanisms described here — individually and in conjunction with one another — to push the frontier of physics into unexplored territory. If cold fusion is conceivable, the scientific community has a societal obligation to prove beyond reasonable doubt that it is impossible. We have not yet fulfilled this obligation.

## Acknowledgments

Helpful conversations with Peter Hagelstein, Matt Trevithick, Ross Koningstein, Yet-Ming Chiang, Curtis Berlinguette, Jeremy Munday, Matt Lilley, Konrad Czerski, David Nagel, and Jim Gimlett are gratefully acknowledged.

## Author contributions

F.M. outlined and wrote the manuscript. C.H. created figures. All authors reviewed and edited the manuscript.

# Supplementary Information

# Known mechanisms that increase nuclear fusion rates in the solid-state

Florian Metzler[1], Camden Hunt[2,3,4,5], Nicola Galvanetto[1,6]

[1]Massachusetts Institute of Technology, Cambridge, Massachusetts 02139, United States

[2]Department of Chemistry, The University of British Columbia, 2036 Main Mall, Vancouver, British Columbia, V6T 1Z1, Canada.

[3]Stewart Blusson Quantum Matter Institute, The University of British Columbia, 2355 East Mall, Vancouver, British Columbia, V6T 1Z4, Canada.

[4]Department of Chemical and Biological Engineering, The University of British Columbia, 2360 East Mall, Vancouver, British Columbia, V6T 1Z3, Canada.

[5]Canadian Institute for Advanced Research (CIFAR), 661 University Avenue, Toronto, Ontario, M5G 1M1, Canada.

[6]University of Zurich, Winterthurerstrasse 190, 8057 Zurich, Switzerland.

# S1. Introduction

*S1.1 Operative definition of readily observable fusion as used in this article*
Here we derive estimates for fusion rates that correspond to experimental observables that are macroscopically and uncontroversially detectable.

In terms of neutron production, Jones et al. 1989[1] report an observed fusion rate of $10^{-23}$ s$^{-1}$ based on neutron detection of $(4.1 \pm 0.8) \times 10^{-3}$ counts s$^{-1}$ (foreground counts minus background counts) at a neutron detection efficiency of $(1.0\pm0.3)\%$ for a 3 g TiD$_2$ sample comprising about $4 \times 10^{22}$ Ti atoms and about the same number of D pairs.

More generally speaking, we considered 100 mW to be a power level that is readily measurable. For instance, the release of 100 mW in a 3 g metal sample in ambient air at 25°C leads to a temperature increase of about 20°C. Translating to nanoscopic units: 100 mW = 100 mJ s$^{-1}$ = $6.2 \times 10^{11}$ MeV s$^{-1}$. For D-D fusion, 23.8 MeV of energy is released per reaction (considering ground state helium as a product), therefore the above corresponds to $2.6 \times 10^{10}$ fusion reactions per second. A 3 g TiD$_2$ sample comprising about $4 \times 10^{22}$ Ti atoms and approximately the same number of D pairs, then 100 mW of observed power release corresponds to a D-D fusion rate of about $10^{-12}$ s$^{-1}$ per D pair.

The above considerations suggest that D-D fusion rates in the range between $10^{-23}$ s$^{-1}$ to $10^{-12}$ s$^{-1}$ and higher can be considered readily observable—subject to experimental details and specific reaction products. Comparing this range with the spontaneous fusion rate in D$_2$ gas (~$10^{-64}$ s$^{-1}$ per deuteron pair at the molecular distance of 74 pm, as estimated by Koonin and Nauenberg[2]) sets a target for fusion rate enhancement of 40-50 orders of magnitude.

*S1.2 Nomenclature when referring to deuteron-deuteron fusion*
The nomenclature for deuterons differs across scientific communities and different framings of the fusion problem. In nuclear physics, it is customary to refer to a deuteron as *d* and to a deuteron-deuteron fusion reaction as *d+d*. Here, the occurrence of collisions with accelerated deuterons is typically implied. In chemistry, deuterons are referred to as *D*, which form *D$_2$* molecules in the gas phase. In quantum dynamics, a deuteron pair such as in the case of a deuterium molecule can be viewed as a single quantum system with a specific energetic state associated with it (at the atomic and at the nuclear level). Such a system can be referred to as |D$_2$>. In this article, we use the *d* terminology, when a collision framing is explicitly adopted and the |D$_2$> terminology when a quantum dynamics framing is explicitly adopted. In the most general sense, we refer to deuteron-deuteron fusion as D-D fusion, a phrase that encompasses all conceivable framings of the process.

## S2. Atomic physics discussion and calculations

*S2.1 General*

A variety of approaches for performing fusion rate calculations based on a two-body tunneling model are provided by multiple authors[3–15]. Central to all these approaches of fusion rate calculations is the tunneling factor *T* that represents the probability of two fusing nuclei reaching the range of strong force attraction (<10 fm). *T* is also referred to as the barrier penetration factor. Nuclear fusion is then assumed to occur essentially instantaneously. The tunneling factor $T(d,U_e)$ depends on the fusion reactants' proximity *d* as well as the screening potential in their vicinity $U_e$ (Figure 2). The defining difference between the received approaches is the choice of interatomic potential *V*. The interatomic potential can be approached as a numerical potential[2,16] or as a parameterized potential[17,18] for numerical evaluation of the integral that determines *T*. Analytical expressions of the integral that serve as approximations are given by some authors[14,15,17].

The tunneling factor *T* can be multiplied by a frequency *f*—which represents the number of attempted fusion events as a function of time—to obtain a fusion rate [19]. The fusion rate obtained in this manner must be subsequently corrected for geometric and nuclear considerations specific to the fusion reaction under consideration (*e.g.,  p+d, d+d, d+t*). This latter correction is canonically known as the S-factor and must be extrapolated from experimental fusion rate data[2,20,21]. However, in conventional beam experiments, no fusion products are observed within reasonable time frames at projectile energies below 1 keV. Accordingly, values of the S-factor at low energies are extrapolated from high energy data and therefore open to debate[22].

All constants can be folded into a single pre-factor *A* that is a physical correction to the bare tunneling probability *T*. When considering the effects of changes in proximity and screening—from an atomic physics perspective, as is the focus of this section—only the factor $T(d,U_e)$ is affected. Accordingly, the discussion here focuses on relative changes of *T* as a function of proximity and screening, and refers to external literature for detailed discussions of pre-factors such as the most appropriate value of the S-factor.

We provide a hosted Python Notebook (Notebook S1) to showcase the essential steps in the calculation of the tunneling probability $T(d,U_e)$:

https://colab.research.google.com/drive/1pEk149hIuTUcNnqN5v6m5FbwtvS1ZpQq

We numerically solve the integral of the tunneling probability expression via the Python function scipy.integrate.quad from the Fortran library QUADPACK.

*S2.2 Deuteron proximity*

Evaluating the Gamow factor for different proximities of deuterons using the methodology described above—and assuming a trial frequency of $10^{15}$ s$^{-1}$ as suggested by previous authors[2,19,23] and assuming a D$_2$ screening energy of 25 eV as suggested by Raiola[24]—yields the following fusion rates for a pair of deuterons:

| Occupancy | Deuteron proximity (pm) | Fusion rate (s$^{-1}$) ($U_e$ = 0 eV) | Fusion rate (s$^{-1}$) ($U_e$ = 25 eV) |
|---|---|---|---|
| O-site | 250 | $10^{-134}$ | $10^{-78}$ |
| O-site/T-site | 170 | $10^{-129}$ | $10^{-75}$ |
| D$_2$/Vacancy | 74–100 | $10^{-98}$-$10^{-114}$ | $10^{-65}$-$10^{-69}$ |

For D$_2$/vacancy occupancy, a proximity range of 74 pm to 100 pm is given. This range is due to the lack of consensus in the literature as to whether deuterium molecules (74 pm) or more distant di-deuterium formations (~100 pm) form in vacancies[25,26].

*S2.3 Electron screening*

In a first approximation, electron screening of nuclei can be expressed as a correction factor ($e^{-r/a}$) applied to the Coulomb potential ($V(r)$), where $a$ is the screening length[27–29]. The screening length in turn consists of a constant multiplied by $n^{-1/6}$, where $n$ is the free electron density of the metal. For palladium, $n$ is approximately 3.15 e$^-$/Å[19]. The resulting screened Coulomb potential is now expressed as:

$$V(r) = \frac{Z_1 Z_2 \, q_e^2}{r} \, e^{\frac{-r}{a}} \tag{Eq. S1}$$

This approach requires the assumption that the free electron density of the metal is treated as a Fermi gas[30]. Application of this approach to a hypothetical system consisting of a D$_2$ molecule within a palladium vacancy, an interatomic distance is contracted from 74 pm to 57 pm[31]. The screening length ($a$) is typically much smaller than the interatomic distance ($r$). Accordingly, the correction factor can be simplified to a subtraction by a constant $U_e$ where $U_e = e^{2/a}$:

$$V(r) = \frac{Z_1 Z_2 \, q_e^2}{r} - U_e \tag{Eq. S2}$$

Other considerations—such as the impact of positive ions on the screening potential as well as dynamic effects—can be introduced to the Coulomb potential as additional corrections[32]. For the purpose of this discussion, they will be considered insignificant and fusion rates will be considered through the context of a constant screening potential $U_e$, as introduced above.

A more detailed expression for the screening potential is given in Czerski et al. 2016:

$$U_e = e^2 k_{TF} = \frac{2e^3}{\hbar \pi^{1/3}} (3\pi n)^{1/6} \sqrt{m^*}$$ (Eq. S2b)

where $k_{TF}$ is the Thomas-Fermi wave number, n the electron density, and $m^*$ the effective electron mass. In first approximation, the effective electron mass is the electron rest mass $m_e$. However, in some solid-state environments, the effective electron rest mass is higher due to local electronic band structure changes that result from particular geometries[33–36].

The two electrons of a gas phase hydrogen/deuterium molecule correspond to a screening potential $U_e$ of ~25 eV[24]. Theoretical $U_e$ values range from 50 to 150 eV for different metals, with Li at the lower range and Pd at the upper range (Figure S1)[28]. The impact of screening energies across this range on the calculated fusion rate is shown in Figure 2. While theoretically predicted screening energies have been calculated for a variety of metals, it should be noted that screening energies obtained by fitting observed fusion rates from experimental data are higher than predicted values[37]. Experiments to determine screening energies have been carried out by multiple groups[38–40] and typically involve low energy deuteron bombardment of different metal targets with concurrent measurement of resulting nuclear byproducts (e.g., neutrons and charged particles). For the same materials where the theoretical screening energy range is given as 50 to 150 eV, the experimental screening energy range reported is 150 to 300 eV (Figure S1)[28,41]. In other words, the experimentally observed fusion rates are substantially higher than expected when only considering proximity and screening within the Gamow model. The typical approach in the literature is to parameterize such discrepancies and include them in the phenomenological correction factor $A$ without understanding them causally.

*S2.4 Time-dependent deuteron proximity and electron screening*
The discussions above on deuteron proximity and electron screening assume that the system is static. Fluctuations are entered in the form of the trial frequency as the rate at which the tunneling probability needs to be compounded. However, dynamic effects and associated temporary increases of proximity and local electron density may affect fusion rates. A key to that argument is that fusion reactions are expected to take place within a timescale of <<1 fs whereas electron oscillations take place at a time scale of >1 fs. Accordingly, even a short extremum in position and electron density would be effectively permanent from the perspective of two fusing nuclei. In other words, such dynamics would be adiabatic in relation to fusion.

Instead of considering the tunneling probability at a single proximity between two nuclei, it has been suggested that integration across all the occurring proximities as nuclei fluctuate is more appropriate[42]. This approach exhibits parallels with the standard practice in thermonuclear fusion of integrating across the full distribution of expected temperatures rather than a single temperature. Subsequent consideration must be given to the magnitude and extent of fluctuations that can be reasonably expected for deuterons in a solid-state material. In a first approximation, fluctuations on the order of 0.1 × the Bohr radius (~5 pm for $D_2$) are conceivable, which would result in an increase in the static D-D fusion rate of ~8 orders of magnitude[42]. Some authors have suggested that such dynamics—and the lack of their consideration—are a primary factor in the discrepancy between theoretical and experimental screening energies[39].

**S3. Nuclear physics discussion and calculations**

*S3.1 General*

The concept of resonance applies both to observable amplitudes in classical systems and to probability amplitudes in quantum systems. In a quantum tunneling problem (*i.e.*, the Gamow model), the probability amplitude beyond the barrier increases when the frequency of the incoming wave (*e.g.*, the energy of an incoming deuteron) matches an intrinsic mode of a nucleus that can result from the interaction. This is illustrated in general terms in Figure 3b where an incident nucleus with energy $E_{incoming}$ is shown on the right and the intrinsic energy levels of the resulting nuclear structure (referred to as compound nucleus) are shown in the center. On the left, cross sections (*i.e.*, tunneling probabilities) are shown as a function of energy. Peaks in tunneling probabilities are seen for incident nucleus energies that match the intrinsic energy levels of the compound nucleus. Such peaks are associated with the concept of resonance.

Whereas Figure 3b is illustrative, Figure 3a shows actual cross sections for common fusion reactions obtained from experimental data. In the given energy range above 5 keV, the d+d fusion reaction exhibits no resonance peak. The D+T fusion reaction exhibits a broad resonance peak centered around 90 keV. The p+$^{11}$B fusion reaction exhibits narrower resonance peaks, for instance a particularly narrow one near 150 keV. The theoretical explanation and prediction of such resonance peaks is the subject of ongoing research in nuclear physics. As alluded to above, the peaks are ultimately a function of the nuclear structure that can form from the interaction of the incoming nucleus with the target nucleus (except if scattering is dominant, as is the case in some configurations). In the d+d case, the resulting structure comprises a four-nucleon system, in the D+T case a five-nucleon system, and in the p+$^{11}$B case, a twelve-nucleon system.

*S3.2 Predicting and measuring nuclear resonances*

Nuclear resonances have traditionally been determined phenomenologically, based on scattering experiments. Alongside such experimental efforts, models have been developed. The models were initially heavily reliant on experimental data but are increasingly based on first principles approaches.

First principles modeling of nuclear reactions draws on nuclear structure models which in turn rely on appropriate nucleon-nucleon interaction models. Such models have been developed based on experimental data that allow for inferences about sizes and shapes of nuclei, energy levels and binding energies, scattering behavior, nuclear reactions and resonances. The top-down deconstruction of phenomenological data then informs the development of bottom-up theoretical models. These models in turn can be used for predictions of new observations in experimentally-inaccessible regimes. In this process of deconstruction and reconstruction, critical assumptions need to be made as to what aspects to include in models and what reductions can be justified in order to keep the resulting models mathematically- and computationally-tractable.

In first approximation toward the development of nuclear models and related intuition, nuclear excited states that cause resonances can be thought of as vibrational modes of nuclei: nuclei can "quiver, ring or even breathe" as Bertsch describes multi-nucleon systems[43]. The modes are caused by nucleon-nucleon interactions and—to a lesser degree—by Coulomb interactions between the positively charged nucleons (*i.e.*, protons). Moreover, nucleons can form molecule-like clusters which further impact the vibrational modes and thus the resulting excited states and resonances. Specifically, the $^4$He compound nucleus that results from D+D fusion is expected to be able to exist in several different 2-2 and 3-1 cluster configurations. Instabilities in the excited states of nuclei lead to different reaction products (decay channels) and angular distributions of reaction products which can be experimentally measured. For instance, the 20.21 MeV state of $^4$He is believed to have a 3-1 structure, where the individual nucleon can be either a proton or a neutron that gets emitted during decay[44].

Nuclear excited states are typically short-lived and excitation occurs in the context of nuclear reactions. Incoming projectiles can transfer energy as well as additional nucleons to the target system. A light nuclear particle (*e.g.*, a photon, neutron, or deuteron) can brush over a target nucleus as a particle wave. If the projectile energy and structure resonate with the target nucleus, nuclear reactions such as absorption can occur rather than elastic scattering. Such absorption would manifest as resonance peaks in cross section diagrams. This is, in essence, the system that needs to be modeled to predict nuclear resonances.

The underlying nucleon-nucleon interaction—also known as the nuclear force or strong force—exhibits a number of features that make it particularly difficult to model: it exhibits three-nucleon effects in addition to two-nucleon effects; it can saturate based on the number of affected nucleons; it is short-range and strongly attractive at <2 fm, yet strongly repulsive at <1 fm; and it is not organized around a center as in the case of electrons around a nucleus.

A central question has been whether nucleon-nucleon interactions could only be explained based on even more fundamental quantum chromodynamics (QCD) models that consider detailed quark interactions of which nucleons are composed, or whether it could be simplified and parameterized to remain at the level of nucleons[45]. In recent years, a wide consensus emerged around chiral effective field theory-based models of the nucleon-nucleon interaction which assume that the interaction is mediated by the exchange of virtual mesons, analog to how the electromagnetic interaction at the atomic scale can be understood as being mediated by the exchange of virtual photons[46]. This approach does exhibit the advantage of representing nucleon-nucleon interactions comparatively accurately at the nucleon level without requiring explicit QCD treatment.

A basic nuclear structure model is the nuclear shell model. The shell model emerges from solving the Schrödinger equation in a mean field nuclear potential such as the Woods-Saxon potential. In this context, "mean field potential" means a potential where all nucleons are assumed to be equally affected by a mean nucleon-nucleon interaction. As can be seen from the listing of characteristics of the nucleon-nucleon interaction in the previous paragraph, this assumption will likely leave out some dynamics among nucleons. Nevertheless, this approach leads to a simple quantum mechanical model—not unlike the harmonic oscillator model for excited states in atoms—that can be solved and that leads to a range of nuclear excited states. Over time, the nuclear shell model evolved to make better use of additional insights gained about nucleon-nucleon interactions. A recent first principles variant of the shell model is the so-called no-core shell model (NCSM) which includes an explicit treatment of three-nucleon interactions [47]. This allows for an integration of the previously discussed chiral effective field theory-based models of the nucleon-nucleon interaction into higher level nuclear structure models. Overviews of recent first principles unification efforts in that direction have been given by Bacca [48] and Quaglioni [49].

The discussion above centered on models for predicting nuclear structure, not nuclear reactions. A reasonable assumption would be that nuclear structure models and nuclear reaction models are closely related. This assumption is—perhaps surprisingly—unfounded. Canonically, nuclear reaction models have not been constructed from first principles but from phenomenological observations and were dependent on experimental input. Meissner summarizes the process of predicting fusion cross sections[50]: "typically, scientists perform experiments at the lowest energy at which fusion reactions can be observed—from thousands down to hundreds of kiloelectronvolts—and then make theoretical extrapolations to lower energies of interest. However, the resulting estimated low-energy data may be unreliable because nucleon dynamics are disregarded in those calculations."

A common method for such extrapolations is the phenomenological R-matrix method[51,52] which provides a framework for fitting experimental data and deriving parameterized cross section estimates. However, even this process is not purely deterministic and requires some judgment, and in some cases "arbitrary" parameter choices, on the side of the researcher[53]. At the heart of the R-matrix method is a separation of the configuration space into an outer region where short-range forces are ignored and an inner region which is considered confined. This approach allows for the calculation of both scattering states (outer region) and bound states (inner region). The R-matrix method has been evolved into a variant known as the computational R-matrix method which is attributed the advantage "that narrow resonances which can escape a purely numerical treatment are easily studied"[53]. In either case, once an R-matrix is calculated for a given nuclear system, then the so-called S-matrix (scattering matrix) can be derived from it. The S-matrix relates initial states to final states and some of its poles are indicative of resonances (and others of bound states). Many of the cross section plots for fusion reactions, as shown in Figure 3a, are determined this way—some relying more on experimental data and others more on computational estimates (and it is not always explicitly stated which is which).

The concept of the S-matrix was first introduced by Wheeler (1937) in the context of developing the Resonating Group Method (RGM). The RGM represents early efforts to link nuclear reaction properties to nuclear structure properties explicitly and is still used today. Originally devised for describing resonant transfer of groups of electrons in scattering processes, the approach was later extended to also apply to groups of nucleons in scattering processes (Nielsen 2016). While the original RGM approach does not consider the structure of nuclei in as much detail as most of the nuclear structure models discussed above, it does consider different clusters of nucleons and the interactions among such clusters. Recent efforts seek to integrate RGM approaches to nuclear reaction modeling and advanced shell model approaches to nuclear structure modeling such as NCSM (see Quaglioni et al. 2012 for a discussion of such efforts). Since NCSM already represents an integration between nucleon-nucleon interaction models and nuclear structure models, such efforts promise to provide a unified picture connecting nucleon interactions to nuclear structure theory as well as to nuclear reaction theory. Models such NCSM will be necessary to predict nuclear resonances rather than experimentally detecting them.

Returning to the concrete case of the d+d reaction, for which fusion rate discrepancies between experimental data and theoretical predictions exist, Czerski et al. ground their proposal for a new resonance near a hypothetical 23.85 MeV excited state of the $^4$He compound nucleus based on RGM calculations[54]. Specifically, they refer to Kanada et al.[55] who suggest that a resonance "very close to the threshold of the d+d channel" may exist[56]. At the same time, the authors imply—like others[44]—that the four-nucleon system is complicated and requires certain assumptions to make such calculations. More research is needed to assert or reject theoretical predictions of proposed near-threshold resonances in the $^4$He system.

All nuclear structure modeling approaches introduced above have a tradeoff between theoretical comprehensiveness and computational tractability. This compromise is expressed by Quaglioni—an expert in NCSM techniques—as follows: "Our model can often contain billions of terms. While more terms improve the accuracy of the model, they also make solving the Schrödinger equation more difficult." Consequently, "these results cannot be considered conclusive until more accurate calculations using a complete nuclear interaction […] are performed."[50] Similarly, Hassid comments on nuclear DFT approaches: "Density functional theory (DFT) is unique in providing a global theory of nuclei. However, it can miss important correlations beyond the mean field."[57] Accordingly, nuclear reaction and resonance prediction is extremely sensitive to small changes in corresponding models. Assumptions made during model development can result in major differences in predicted outcomes. Quaglioni emphasizes that predicted resonances are "extremely sensitive to higher-order effects in the nuclear interaction, such as three-nucleon force (not yet included in the calculation) and missing isospin-breaking effects in the integration kernels."[50] While nuclear resonances are still explored computationally[58,59], the manifestation of the challenges above is that nuclear resonances continue to be primarily confirmed through experiment. Ultimately, our knowledge of nuclear processes is confined by the empirical limits of engineering.

*S3.3 Calculations showing the impact on fusion rates of a hypothetical DD resonance at 23.84 eV*

We expand upon the D-D fusion rates calculations for the fusion rate reported in Figure 2 by overlaying the corresponding cross section with a hypothetical resonance centered at 23.84 eV, as proposed by Czerski and co-workers[33,34]. Following these authors, the impact of the hypothesized resonance on the cross section is represented by a Breit-Wigner distribution of the form:

$$\sigma_{res}(E_{cm}) = \frac{\pi}{k^2} \frac{\Gamma_d \Gamma_p}{(E_{cm} - E_R)^2 + \frac{1}{4}\Gamma^2}$$ (Eq. S3)

whereby $k$ is the Thomas-Fermi wave number, $\Gamma_d$ is the partial deuteron width, $\Gamma_p$ is the partial proton width, $E_{cm}$ is the relative center-of-mass energy between the deuterons, $E_R$ is the resonance energy, and $\Gamma$ is the resonance width. Note that with its dependence on $k$ such a resonance – although originating from nuclear dynamics – would vary as a result of local lattice conditions. Figure S2 illustrates the impact of a nuclear resonance with $E_R$ = 1 eV and $\Gamma$ = 0.5 eV on a D-D fusion reaction cross section. If such a resonance can indeed be shown to exist, then the precise parameters of the distribution that represent the impact of the resonance (such as a scaling factor) are best determined experimentally.

We provide a hosted Python Notebook (Notebook S2) to showcase the essential steps in the calculation of this resonance-based enhancement $\sigma_{res}(E_{cm})$ to the overall fusion cross section:

https://colab.research.google.com/drive/183HIJucKdrRP8Ep9otROiRRUYcHn_aJB?usp=sharing

### S4. Quantum dynamics calculations

*S4.1. General*
Nuclear fusion can be modeled in approximation as a two-level system (TLS) undergoing spontaneous emission at a rate corresponding to the fusion rate: an excited TLS (i.e., the fusion reactants) relaxes to its ground state (i.e., the fusion products)[60].

If the number of final states is large enough to be represented as a continuum of states, the so-called Golden Rule can be applied to extract a rate, as shown below[61]. That is indeed the case with conventional (incoherent) fusion where the possible final states represent a continuum. The equivalence of modeling tunneling via the Wentzel–Kramers–Brillouin (WKB) approximation (as shown in section S1) and modeling spontaneous emission via the Golden Rule (as shown here) is explicated by Raju[62].

The simplest Hamiltonian representation of a TLS ($H_{TLS}$) that relaxes from an excited state (*i*) to its ground state (*f*) is:

$$H_{TLS} = \begin{bmatrix} i & V \\ V & f \end{bmatrix}$$ (Eq. S3)

where $V$ represents the coupling between the states $i$ and $f$. A corresponding diagram is shown in Fig. S3. The described treatment of the D-D system as a two-level quantum system is discussed in greater depth by Hagelstein[63,64].

Now consider the $^4$He + γ channel of D-D fusion: the photon can be emitted in many unique angular configurations and the photon can take on a distribution of energies near 23.84 MeV. Accordingly, the number of final states are effectively infinite and lead to irreversible dynamics[65]. In this case, the Hamiltonian matrix more accurately is:

$$H_{GoldenRule} = \begin{bmatrix} i & V & V & V & V & V & \dots \\ V & f_1 & 0 & 0 & 0 & 0 & \dots \\ V & 0 & f_2 & 0 & 0 & 0 & \dots \\ V & 0 & 0 & f_3 & 0 & 0 & \dots \\ V & 0 & 0 & 0 & f_4 & 0 & \dots \\ V & 0 & 0 & 0 & 0 & f_5 & \dots \\ \vdots & \vdots & \vdots & \vdots & \vdots & \vdots & \ddots \end{bmatrix}$$

(Eq. S4)

The transition rate (Γ), per the Golden Rule, from an initial state $i$ to a continuum of final states $f$ is[61]:

$$\Gamma_{i \to f} = \frac{2\pi}{\hbar} |\langle f | H | i \rangle|^2$$

(Eq. S5)

With the latter term taking on the value of:

$$\langle f | H | i \rangle = V = e^{-G}$$

(Eq. S6)

where the transition matrix element $V$ between $i$ and $f$ is represented by the inverse exponential of the Gamow factor ($G$), which represents the penetration probability. The small value of $V$ represents the low probability of the nuclear transition. This treatment results in an equivalency between the transition rate and the well-known fusion rate (section S1.2) for a two-nuclei system:

$$\Gamma_{i \to f} = e^{-2G}$$

(Eq. S7)

$\Gamma_{i \to f}$ represents a decay channel that is always available at the given rate in a two-nuclei system. However, it is not necessarily the only decay channel. If a coupling to a resonant/near-resonant quantum system exists, additional dynamics must be considered. Specifically, in the case of precise resonance with an acceptor system, a superposition state results in energy being distributed across both donor and acceptor systems and in an occupation probability that oscillates back and forth with a coupling-dependent Rabi frequency $f_{Rabi}$, which is given as[66]:

$$\Omega_{i,f_{Rabi}} = \sqrt{|\Omega_{i,f_{Rabi}}|^2 + \Delta^2} \qquad (Eq.\ S8)$$

where the detuning value ($\Delta$):

$$\Delta = \omega_{\text{transition1}} - \omega_{\text{transition2}} \qquad (Eq.\ S9)$$

vanishes in the resonance case.

The Hamiltonian now corresponds to the quantum Rabi model[67]:

$$H_{Rabi} = \overbrace{\hbar\omega_0\sigma_z}^{tls} + \overbrace{\hbar\omega a^\dagger a}^{osc} + \overbrace{\hbar g\left(a^\dagger + a\right)\sigma_x}^{interaction} \qquad (Eq.\ S10)$$

where $\hbar\omega_0$ is the transition energy in the TLS, $\hbar\omega$ is the energy in the oscillator mode, and $\hbar g$ is the energy in the interaction between them. $\sigma_x$, $\sigma_y$, $\sigma_z$ are Pauli matrices and $a^\dagger$ and $a$ are creation and annihilation operators. Time-evolution of this system yields an occupation probability that Rabi oscillates between the TLS and the cavity (Figure S5). While the above implementation illustrates what a coherent D-D fusion channel could look like, no cavity can be practically implemented at this time that can readily absorb the 23.8 MeV energy that would result from the $|D_2\rangle \rightarrow |^4He\rangle$ reaction.

However, a modified version of such a system can be implemented, where another TLS (such as a nucleus) with a resonant excited state absorbs the released energy and where a bosonic mode mediates the transfer (Figure S6). In the case of non-radiative transfer, the mediating bosonic mode can facilitate the transfer of a large quantum of excitation (in our case 23.8 MeV) without ever having to hold the entirety of that energy itself[68–70]. The coupling between two nuclei in that case is indirect, i.e. via the mediating bosonic mode, whereas the coupling from each nucleus to the bosonic mode is direct.

Nuclei able to participate in such mediated transfer as acceptor systems would need to exhibit a resonant excited state at the $|D_2\rangle \rightarrow |^4He\rangle$ transition energy of 23.8 MeV. A $^4He$ nucleus would be precisely resonant as it can undergo the same reaction in reverse. However, in this case a $|D_2\rangle \rightarrow |^4He\rangle$ reaction would be offset by a $|^4He\rangle \rightarrow |D_2\rangle$ reaction within a closed system. Accordingly, the process would be essentially impossible to observe. Alternatively, if the

acceptor nuclei were to incoherently disintegrate upon receipt of the transferred excitation—thus turning the closed quantum system into an open one—the process would be observable. Many heavy nuclei can resonantly absorb near 23.8 MeV and would promptly, incoherently decay via alpha, proton, or neutron emission upon receipt of such a large quantum of excitation[71]. Experimental investigations of this mechanism for D-D fusion would require nuclei that exhibit dense states near 23.8 MeV that are stable enough for many nuclei to participate in collective coherent dynamics, yet unstable enough such that disintegration occurs at an observable rate.

Building upon these criteria for suitable acceptor nuclei, we assume that the acceptor nuclei is stable within the relevant timescale and has a high density of states near 23.8 MeV. The Hamiltonian for a system containing such nuclei with resonant states now corresponds to a Dicke model:

$$H_{Dicke} = \hbar\omega_0 \sum_{j=1}^{N} \sigma_z^j + \hbar\omega a^\dagger a + \hbar g \sum_j \sigma_x^j (a + a^\dagger)$$
(Eq. S11)

where index $j$ counts over $N$ nuclei and $N$ corresponding interaction terms with the shared oscillator mode.

Similar to the previous case (Eq. S10) time-evolution of this system yields an occupation probability that Rabi oscillates between the TLSs, mediated by the cavity (Figure S6). Note that in the above example, the coupling as well as the energy in the mediating oscillator are small in relation to the state transition energy. Comparable dynamics result even if the coupling and the energy in the mediating oscillator are very small in relation to the state transition.

We demonstrate classical toy models that exhibit analogous excitation transfer dynamics in a hosted Python notebook (Notebook S3):

https://colab.research.google.com/drive/185enSM5hF2oq-NHhEu8yjGE1y6OAWQoh

*S4.2 Expanding models from two-level states to nuclei*
Moving the modeling process from generic two-level systems and couplings to the concrete case of atomic nuclei coupled to a shared oscillating magnetic field ($\hbar\omega$ with interaction $-\mu \cdot B$), the Hamiltonian becomes[63,64]:

$$H_{Nuclear} = \overbrace{\sum_j M_j c^2}^{nuclei} + \overbrace{\hbar\omega a^\dagger a}^{osc} + \overbrace{\sum_j -\mu_j \cdot B}^{magn.\,interaction}$$
(Eq. S12)

with $M_j$ containing two nuclear states:

$$M_j = \begin{bmatrix} M_1 c^2 & 0 \\ 0 & M_2 c^2 \end{bmatrix}$$
(Eq. S13)

An example for a nuclear transition interacting with an oscillating magnetic field is the dipole M1 transition[72]. To obtain a rate equation for this resonance energy transfer mechanism requires determining the matrix element that describes the indirect coupling through intermediate states, as shown in Figure S7.

Hagelstein 2018[63] provides in Eq. 64 the indirect coupling matrix element that drives nuclear excitation transfer in the general case. Below, that coupling matrix element is evaluated for the concrete case of the $D_2 \rightarrow {}^4He$ transition:

$$\left\langle \Psi_{1'} \Psi_{osc} \middle| \hat{H}_{int} \left( E - \hat{H}_0 \right)^{-1} \hat{H}_{int} \middle| \Psi_1 \Psi_{osc} \right\rangle \rightarrow \left[ \frac{[g e^{-G} \sqrt{\frac{vol_{nuc}}{vol_{mol}}}] g}{\Delta E} \right] \left[ \frac{\Delta E_+ - \Delta E_-}{\Delta E} \right] \left[ \sqrt{N_{D_2}} \sqrt{N_{He}} \right]$$

(Eq. S14)

Here the first factor describes basic excitation transfer dynamics, the second factor accounts for nuclear off-shell effects, which are on the order of 1 (as discussed in detail by Hagelstein[63,64]), and the third factor describes Dicke enhancement as discussed further below.

The first term represents the transition from the initial $|D_2\text{ He-4}\rangle$ state to a virtual state, where the $D_2$ molecule in the system shifts to a compact four-nucleon configuration (i.e. He-4). This transition is driven by the interaction with another nucleus via the coupling of each nucleus to the bosonic mode $g$ and is impeded by a hindrance factor comprising a Coulomb barrier factor $e^{-G}$ and a volumetric factor $\sqrt{vol_{nuc}/vol_{mol}}$ [73]. The Coulomb barrier factor represents the penetration probability of the Coulomb barrier that needs to be tunneled through to reach the compact He-4 state. The volumetric factor accounts for the volumetric change from a molecular-scale four-nucleon configuration to a nuclear-scale four-nucleon configuration. It can be broken down into:

$$\frac{vol_{nuc}}{vol_{mol}} = \frac{\frac{4}{3}\pi r_{nuc}^3}{2\pi^2 R_0 \Delta R^2} = 6.26 \times 10^{-12}$$
(Eq. S16)

where $r_{nuc}$ represents the nuclear radius and $R_0$ and $\Delta R$ represent parameters of the radial molecular $D_2$ wavefunctions[73].

The product of the coupling constant $g$ and the hindrance factor represents the probability of entering a virtual state. The second $g$ along a transfer path in Figure S7 represents the probability of emerging from the virtual state to the final state.

Note that the Coulomb factor and the volumetric factor come into the rate equation only once. That is because from the perspective of nuclear states, a separation of two deuteron pairs on the order of several femtometer is sufficient to absorb energy in the form of additional nuclear binding energy – in this case, the 23.8 MeV received via resonance energy transfer that adds to the mass of the separated two-nucleon clusters (i.e. two deuterons in close proximity) over the more compact four-nucleon configuration (i.e. the ground-state helium nucleus). However, at such short distances, the nuclear force still dominates the electrostatic force, preventing immediate ejection of the newly formed nuclear deuteron clusters. The deuterons can tunnel apart with a given probability—akin to alpha emission via tunneling in heavier nuclei—but because this process is hindered through the outward potential barrier, most newly formed $D_2$ acceptor systems can be expected to initially remain in their compact pre-fission state[74–77]. The clustered two-deuteron pair state $D\text{-}D_{10\ fm}$ has been referred to as $D_2$ compact state[78].

Newly formed $D_2$ compact states can participate in further dynamics within their lifetime. For instance, when coupled to resonant acceptor systems such as ground-state He-4 nuclei, they can act as donors of 23.8 MeV, which they can shed through return to the lower energy helium configuration. The latter is the case with Rabi oscillations, e.g. between coupled $D\text{-}D_{10\ fm}$ and He-4 systems. Such follow-on dynamics are much faster than the initial $|D_2\rangle \rightarrow |\text{He-4}\rangle$ transition, since they are driven by the same couplings but no longer involve the hindrance factor representing Coulomb barrier penetration and volumetric change.

Many nuclei are able to form compact clustered states capable of absorbing energy via nuclear binding without immediately disintegrating[79–85]. Further research on cluster models applied to different nuclides—as discussed in section S2.2—may further elucidate the available states and their properties. Those nuclei whose clustered states are sufficiently long-lived can participate—across significant timescales (e.g. ps to ns range)—in collective dynamics with many participating nuclei, therefore driving up Dicke enhancement factors that accelerate the overall dynamics.

Dicke enhancement is captured in the third term of the coupling matrix element represented by Eq. S14. Dicke enhancement describes the speedup of certain coherent quantum processes as a function of the number of participating systems[86]. These scaling dynamics are described in Abasto et al. 2012 where they are referred to as supertransfer[87].

From the coupling matrix element, the corresponding transfer rate is obtained:

$$\Gamma = \frac{\left\langle \Psi_{1'}\Psi_{osc} \middle| \hat{H}_{int}\left(E - \hat{H}_0\right)^{-1}\hat{H}_{int} \middle| \Psi_1\Psi_{osc}\right\rangle}{\hbar} = \frac{[ge^{-G}\sqrt{\frac{vol_{nuc}}{vol_{mol}}}]g}{\Delta E}\sqrt{N_{D_2}}\sqrt{N_{He}}}{\hbar}$$

(Eq. S17)

This rate expression needs to be evaluated for different available couplings $g$ to determine which coupling drives the dominant channel.

For electric coupling, $g = d \cdot E$ where $d$ is ~e × 1 fm and $E$ is ~$10^5$ V/cm, resulting in a $g$ estimate on the order of $10^{-10}$ eV. This value broadly matches the electric-nuclear couplings given in Engel et al. 2013[88]. For magnetic coupling, $g = -\mu \cdot B$ where $\mu = 3.15 \times 10^{-8}$ $eV$ $T^{-1}$ is the nuclear magneton and $B$ is the internal magnetic field affecting nuclei in a lattice. For $B$ on the order of 3 T, an estimate for $g$ on the order of $10^{-7}$ eV results. For an exemplary measurement of internal magnetic fields in a metal lattice, consult Koi et al. 1961[89]. For relativistic phonon-nuclear coupling, $g = a \cdot cP$ with $a$ estimated[73,90,91] to be between $10^{-6}$ and $10^{-8}$ and a center-of-mass momentum of nuclei $P$[92] on the order of $10^{-24}$ kg m s−1. This results in an estimated coupling $g$ on the order of $10^{-3}$ eV.

Next we evaluate the rate expression (Eq. S17) using a conservative coupling estimate $g = 10^{-7}$ eV, the value given in Eq. S16 as the volumetric factor, and assuming $10^{12}$ $D_2$ donor pairs and $10^6$ He-4 acceptor nuclei in the coherence domain of the shared oscillator mode. This yields the following fusion rate estimate:

$$\Gamma = \frac{10^{-7}eV\, 10^{-33}\, 10^{-6}\, 10^{-7}eV}{24 \times 10^6 eV} 1 \sqrt{10^{12}}\sqrt{10^6}\frac{1}{\hbar} = 10^{-34} s^{-1}$$ (Eq. S18)

Finally, we add 150 eV of screening applied to the Gamow factor $G$, as discussed in section S1:

$$\Gamma = \frac{10^{-7}eV\, 10^{-17}\, 10^{-6}\, 10^{-7}eV}{24 \times 10^6 eV} 1 \sqrt{10^{12}}\sqrt{10^6}\frac{1}{\hbar} = 10^{-18} s^{-1}$$ (Eq. S19)

The result is a predicted fusion rate for $D_2$ in a coherently stimulated metal hydride sample that is enhanced by about 46 orders of magnitude over the estimated rate of $10^{-64}$ $s^{-1}$ for $D_2$ gas at ambient conditions. Note that via the enhancement factor, the coherent fusion rate is strongly dependent on the number of deuteron pairs in a given coherence domain. As a result, both increasing the deuteron density and enlarging the coherence domain lead to further rate increases.

**S5. Earlier reports of energetic particle emission from metal-hydrogen systems at unexpected energies**

A number of articles report observations of energetic particles emitted from stimulated metal-hydrogen systems at energies that do not correspond to canonical nuclear reactions expected to take place in such systems. A selection of such articles and a brief summary of their reported observations are given below:

In Chambers et al. 1990[93], the authors report the observation of charged particles with energies ~28 MeV from Pd foils that were bombarded with deuterium ions at energies <1.5 keV.

In Takahashi et al. 1990[94], the authors report the observation of neutron emission in the 3-7 MeV range from electrochemically loaded PdD samples stimulated by electric current pulses.

In Chambers et al. 1991[95], the authors report the observation of ~5 MeV charged particles from the bombardment of Ti foils with deuterium ions at energies of 350-400 eV.

In Ziehm 2022[96], the author reports the observation of 138 keV charged particles from the bombardment of Pd foils with deuterium ions at energies <500 eV.

**S6. Supplemental Figures**

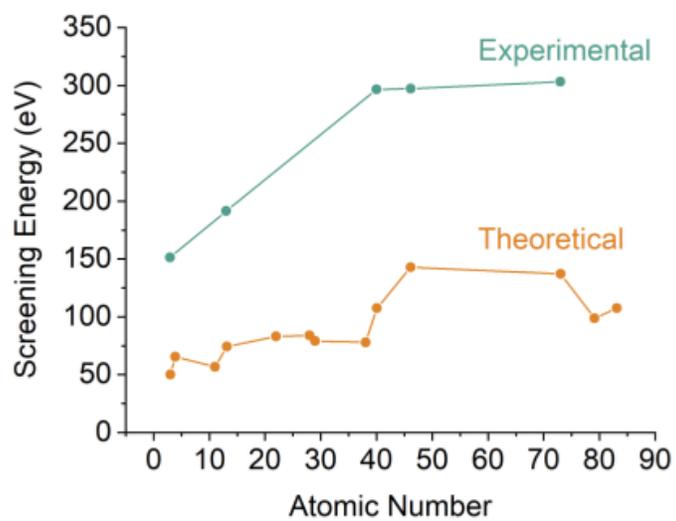

**Figure S1.** Reported experimental (orange) and theoretical (teal) screening energy values. Reproduced with permission from Huke et al. 2008[28].

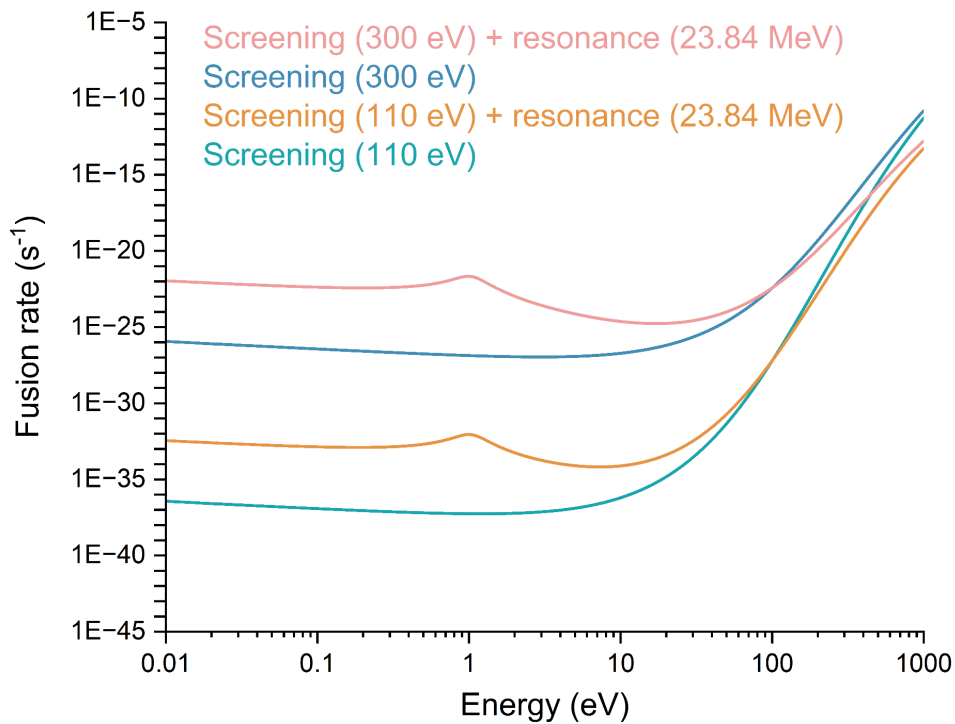

**Figure S2.** d+d fusion rate estimates assuming: i) 110 eV with no d+d resonance; ii) 110 eV with a narrow d+d resonance centered at 23.84 eV; iii) 300 eV with no d+d resonance; iv) 300 eV with a d+d resonance centered at 23.84 eV. Full calculations—based on original work by Czerski and co-workers[33,54]—can be found in Notebook S2.

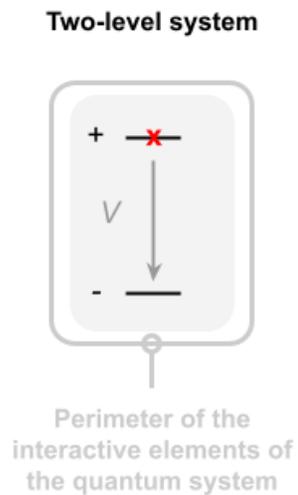

**Figure S3.** Essential features of a two-level system (TLS): An excited state (+) is coupled to a ground state (-) whereby the coupling is represented by $V$. A perimeter encompasses all interacting elements of the quantum system, which in this case is only the TLS itself.

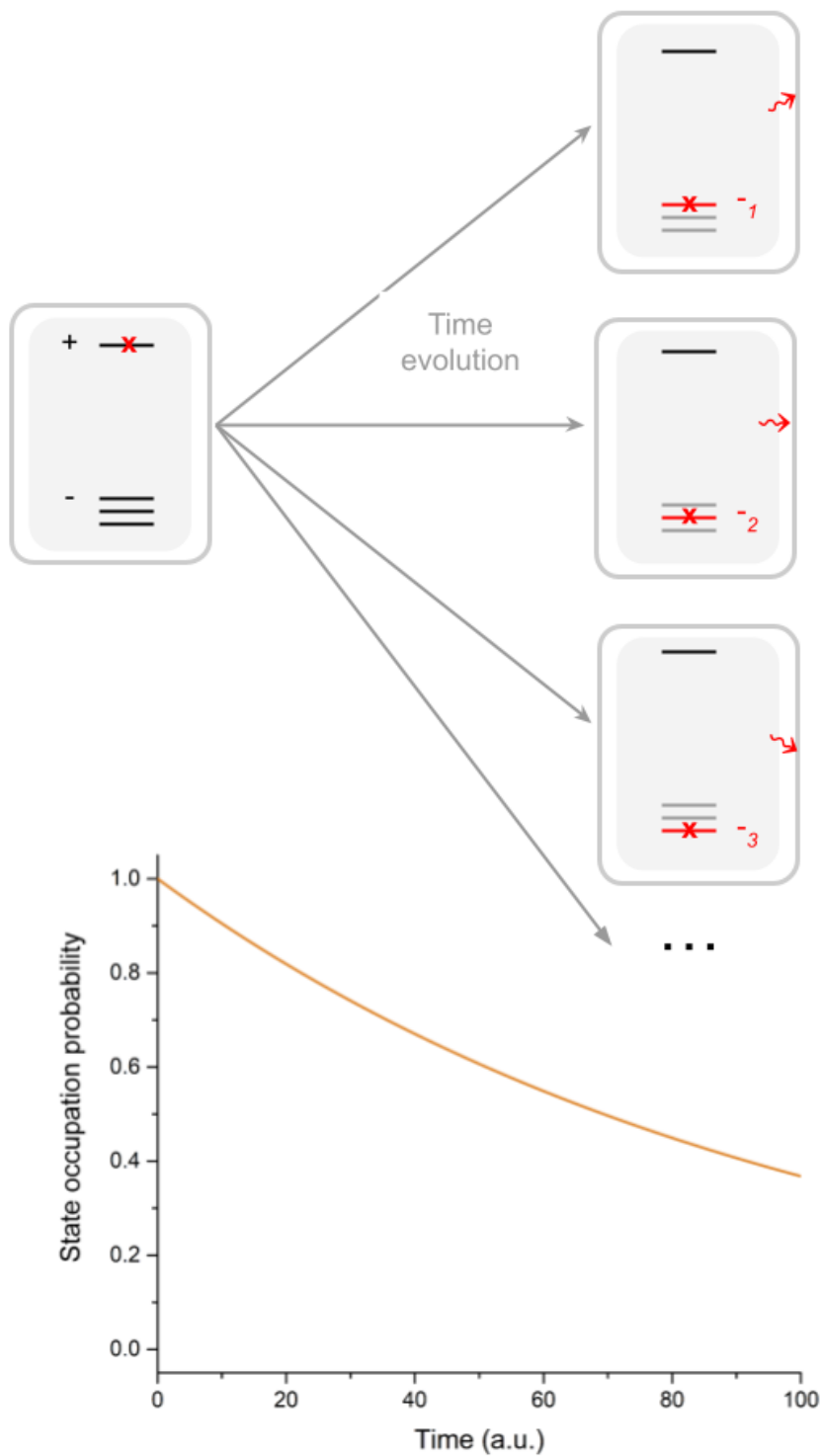

**Figure S4.** A two-level system (TLS) with a continuum of final states: When a TLS can relax into a near-infinite number of final states, the transition probability takes the form of a decaying exponential per the Golden Rule expression (Eq S5). Here, the near-infinite number of final states are represented by small differences in final state energies and in differences in the angular orientation of resulting photon emission.

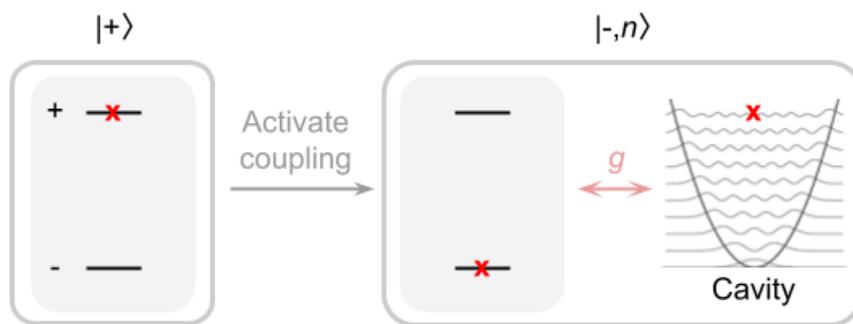

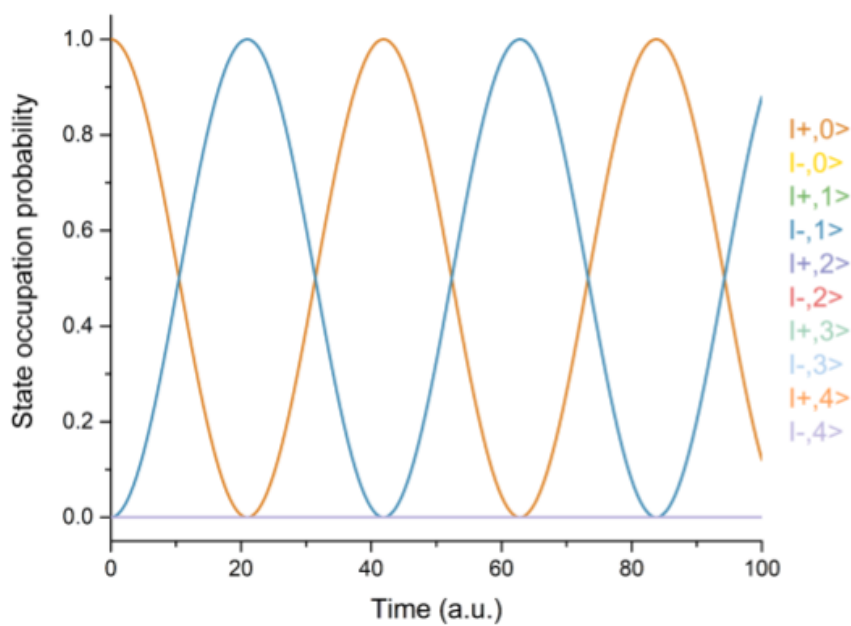

**Figure S5.** A two-level system (TLS) coupled to a cavity with a resonant oscillator mode: the coupled resonant oscillator mode allows for direct energy exchange between the TLS and the cavity, resulting in Rabi oscillations. Simulations of corresponding dynamics based on the quantum dynamics Python library QuTiP are given in the Python notebook hosted at:
https://github.com/project-ida/two-state-quantum-systems/blob/master/03-a-two-state-system-in-a-quantised-field.ipynb

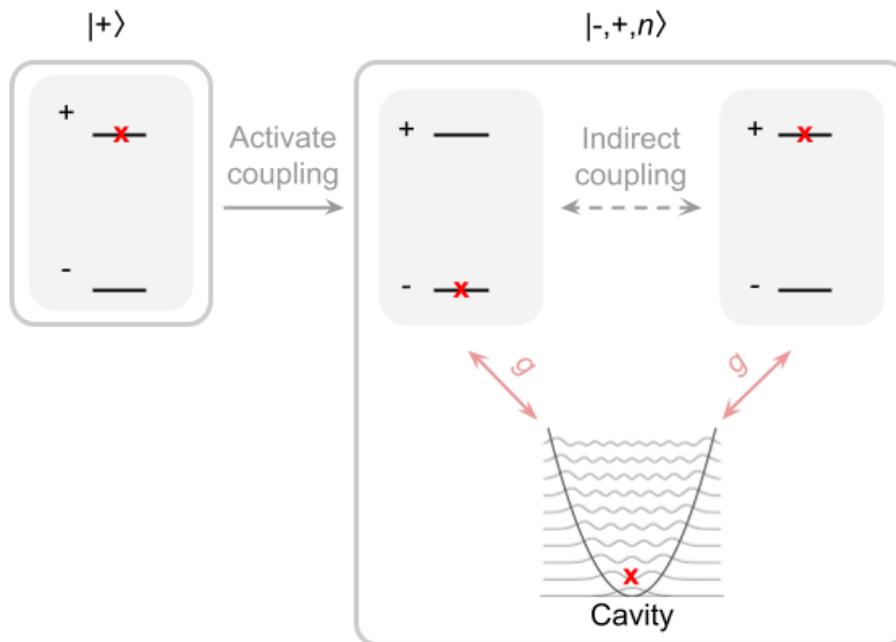

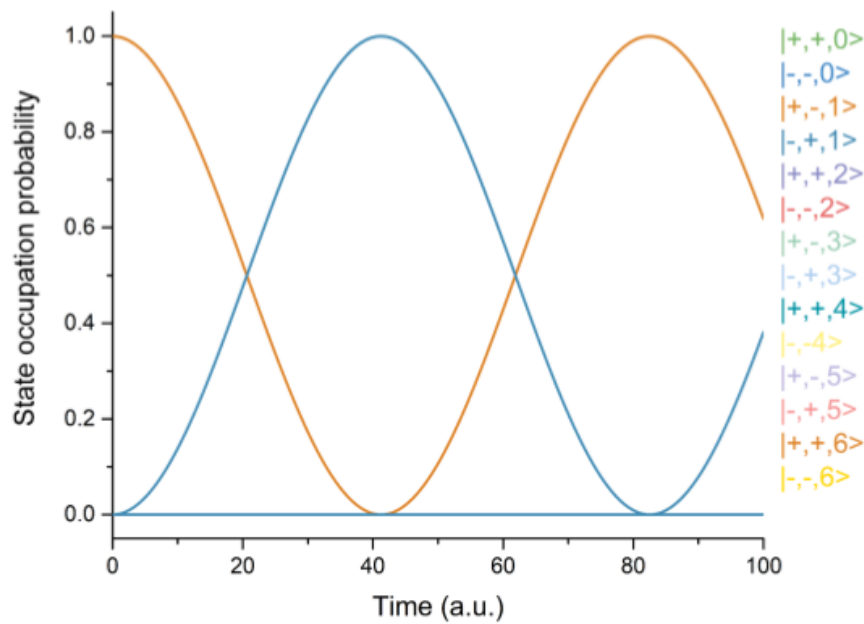

**Figure S6.** Two two-level systems (TLS) coupled to a shared oscillator mode with resonant TLS states: the coupled resonant TLS allows for indirect energy exchange between the two TLSs via the shared oscillator mode, resulting in Rabi oscillations. Simulations of corresponding dynamics based on the quantum dynamics Python library QuTiP are given in the Python notebook hosted at:
https://github.com/project-ida/two-state-quantum-systems/blob/master/05-excitation-transfer.ipynb

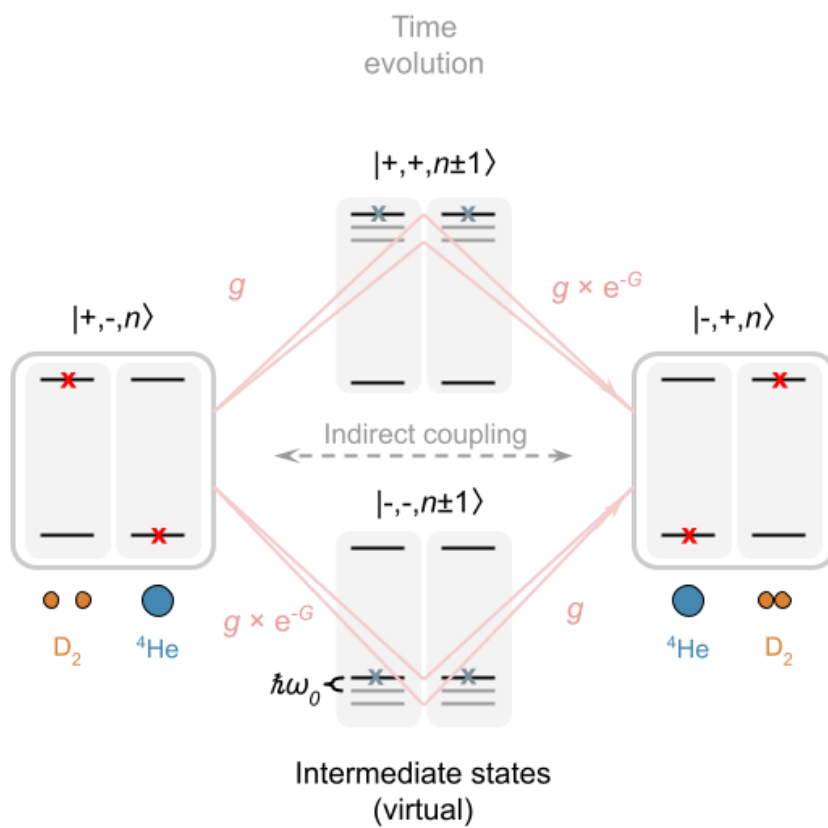

**Figure S7.** Two nuclear systems ($D_2$ and $^4$He) coupled to a shared oscillator mode with resonant nuclear states: resonance energy transfer occurs through the temporary occupation of virtual states, in which either both of none of the two nuclear systems are excited. In the case of D-D fusion, the coupling from a D—D$_{74\text{ pm}}$ to $^4$He nuclear state is impeded by a hindrance factor $e^{-G}$ but not from a $^4$He to a D-D$_{10\text{ fm}}$ nuclear cluster state.